\documentclass{JFM-FLM_Au}

\usepackage{float}
\usepackage{xcolor}
\usepackage{graphicx}
\usepackage{subcaption}
\usepackage{amsmath}

\AtBeginDocument{
	\hypersetup{
		colorlinks=true,
		citecolor=blue,
	}
}

\lefttitle{T.-Y. Han, J.-Y. Pan, T. Long,  E. Cipriano, A. Millischer, M.-J. Ni and S. Zaleski}
\righttitle{Journal of Fluid Mechanics}

\title{Direct numerical simulations of micrometre-scale film dewetting with phase change}

\author{Tianyang Han\aff{1,2}, Jieyun Pan\aff{2}, Tian Long\aff{3}, Edoardo Cipriano\aff{2}, Alexandre Millischer\aff{2}, Mingjiu Ni\aff{1} \and Stéphane Zaleski\aff{2,4}}

\affiliation{\aff{1}School of Engineering Science, University of Chinese Academy of Sciences, 101408, Beijing, China
\aff{2}Sorbonne Université, CNRS, Institut Jean Le Rond d'Alembert, 75005, Paris, France
\aff{3}State Key Laboratory for Strength and Vibration of Mechanical Structures, School of Aerospace, Xi’an Jiaotong University, Xi’an, 710049, Shaanxi, China
\aff{4}Institut Universitaire de France, Paris, France}

\corresau{Jieyun Pan, \email{yi.pan@sorbonne-universite.fr}; Mingjiu Ni, \email{mjni@ucas.ac.cn}; Stéphane Zaleski, \email{stephane.zaleski@sorbonne-universite.fr}}

\begin{document}
\maketitle

\begin{abstract}

Motivated by the microlayer beneath vapour bubbles in nucleate boiling, we perform two-dimensional direct numerical simulations of micrometre-scale film dewetting with phase change on a superheated wall. Finite interfacial heat-transfer resistance and Navier slip regularize the thermal and viscous singularities at the contact line. The isothermal results agree reasonably well with asymptotic predictions for viscous dewetting. With phase change, increasing wall superheat or accommodation coefficient accelerates contact-line recession and increases the apparent contact angle, particularly at small equilibrium contact angles. The interfacial heat flux is concentrated near the contact line and is well described by local one-dimensional conduction. This localized evaporation drives liquid towards the contact line, opposing capillary-driven retraction farther away and producing a near-contact-line region where the interfacial slope increases with distance. Outside this region, the slope distributions cannot be collapsed using a Cox--Voinov-type relation with an evaporation-dependent Voinov angle and length, as assumed in scale-separated descriptions. A scaling analysis of the lubrication equation shows that such a collapse requires films much thicker than the interfacial thermal-resistance length; in the present films, evaporation and contact-line motion compete up to the film-thickness scale. This competition persists at a reduced slip length: contact-line motion weakens the slope increase near the contact line, and evaporation weakens the slope decrease farther away. Accordingly, a lubrication model captures the overall simulated slope distributions only when both contributions are retained.

\end{abstract}


%{\bf MSC Codes }  {\it(Optional)} Please enter your MSC Codes here

\section{Introduction}
\label{sec:Introduction}

The motion of a liquid--vapour--solid contact line over a superheated wall is central to a number of phase-change processes, including sessile-drop evaporation, spray cooling and nucleate boiling. In nucleate boiling, a thin liquid layer with a thickness of a few micrometres may form beneath a growing vapour bubble and can contribute substantially to the heat transfer coefficient at the solid wall \citep{burevs2022comprehensive,tecchio2024microlayer,long2025direct}. The formation of this microlayer is closely related to the motion of the contact line at the bubble foot along the heated substrate: depending on the wall superheat, thermal boundary condition and contact-line speed, heat transfer may occur either mainly in the immediate vicinity of the contact line or through an extended microlayer left behind the moving interface \citep{schweikert2019transition,burevs2021modelling}. From this viewpoint, the dynamics of microlayers near the contact line share important features with film dewetting in the presence of phase change. Understanding how evaporation modifies the motion of a receding contact line (moving into the liquid side) and the shape of a micrometre-scale liquid film is therefore a necessary step towards modelling microlayer dynamics in nucleate boiling.

A striking feature of evaporating contact lines is that the apparent contact angle, measured from the interface shape at a macroscopic scale, can differ substantially from its isothermal value. For highly wetting FC-72 droplets on superheated substrates, \citet{raj2012contact} observed relatively large macroscopic contact angles. Using a lubrication-based model following \citet{stephan1994}, they further showed that the apparent contact angle depends on the wall superheat and the thickness of the adsorbed film ahead of the contact line. More directly related to dewetting, \citet{fourgeaud2016} studied evaporating ethanol films deposited by a receding meniscus and observed the formation of a dewetting ridge, even though the liquid completely wets the substrate at equilibrium. They attributed this behaviour to an evaporation-induced increase in the apparent contact angle, caused by the strong liquid flow towards the highly evaporating region near the contact line and the associated viscous bending of the interface. These observations indicate that evaporation can modify contact-line dynamics not only by removing liquid mass, but also by altering the near-contact-line flow and the resulting interfacial slope.

To distinguish the effects of evaporation from those associated with contact-line motion, it is useful to first consider the limiting case without phase change. Isothermal dewetting of a thin liquid film is a classical moving-contact-line problem in which capillary forces drive the recession of the contact line, while viscous stresses dissipate energy in the liquid wedge and in the growing rim \citep{gennes2004capillarity}. Outside the microscopic slip region, the interface slope is expected to follow the Cox--Voinov logarithmic variation \citep{voinov1976hydrodynamics,Cox1986the}. For a dewetting film, \citet{snoeijer2010asymptotic} used matched asymptotic analysis to connect the microscopic contact-line region to the growing rim and the outer film, thereby deriving expressions for the contact-line speed and apparent contact angle in terms of the equilibrium contact angle, slip length, rim size and initial film thickness. This theory therefore provides a natural baseline against which the additional effects of phase change on the contact-line motion and interface shape can be assessed.

When evaporation is present, lubrication-based models have been developed to describe the coupled heat transfer and viscous flow in the thin liquid wedge near the contact line. In these models, the viscous and thermal singularities at the contact line can be regularized by microscopic effects such as slip, disjoining pressure, the Kelvin effect or a finite interfacial heat-transfer resistance \citep{hocking1995contact,ajaev2010static,janevcek2012contact}. The resulting solutions relate the microscopic contact angle imposed at the solid surface to the apparent contact angle observed at the macroscopic scale. For evaporating contact lines in the absence of contact-line motion, this framework predicts that increasing the wall superheat enhances the evaporation-induced flow towards the contact line and therefore increases the apparent contact angle \citep{hocking1995contact, morris2001contact}. However, when the contact line moves, the interface shape is also affected by the viscous bending associated with contact-line motion. This has motivated a scale-separated description in which the influence of phase change is assumed to be confined primarily to a micro-region near the contact line, as illustrated in figure~\ref{fig:2-1-1}(b). In this description, the influence of evaporation on the outer interface is represented by two effective parameters: the Voinov angle $\theta_V$, towards which the interfacial slope increases within the micro-region, and the Voinov length $l_V$, which serves as an effective microscopic length scale. These parameters then enter the outer Cox--Voinov-type relation governed by contact-line motion \citep{janevcek2013apparent,fourgeaud2016,zhang2022dewetting}.

To determine the Voinov angle in this scale-separated description, lubrication equations are commonly solved within the evaporation-dominated micro-region using a shooting method, in which contact-line motion is usually neglected \citep{raj2012contact,huber2017direct,torres2024coupling}. This simplification allows the interfacial slope to approach a constant value in the far field of the micro-region, corresponding to vanishing curvature, and thus provides a convenient boundary condition for the shooting procedure. Using this approach, \citet{zhang2022dewetting} showed that, for $50\,\mathrm{\mu m}$-thick films, the scale-separated description reproduces the contact-line speed and apparent contact angle predicted by full lubrication solutions retaining both evaporation and contact-line motion. For microlayers only a few micrometres thick, however, the separation of scales underlying this description is not guaranteed. It remains unclear whether contact-line motion can be neglected within the micro-region and whether evaporation-induced changes in the interfacial slope  can be captured solely through the Voinov angle and length.

Investigating these questions in full nucleate-boiling simulations is challenging because of the large separation of spatial scales: the computational domain must accommodate the growth of a millimetre-scale vapour bubble while resolving a micrometre-scale microlayer and an even
finer region near the contact line. Over the last decade, direct numerical simulations (DNS) have been increasingly used to obtain a more detailed description of microlayer dynamics in nucleate boiling. Resolved simulations with micrometre-scale grid spacing have been performed to study bubble growth and microlayer evolution \citep{huber2017direct,urbano2018direct}. More recent studies have achieved minimum grid spacings of approximately $0.5\,\mathrm{\mu m}$ and have further incorporated interfacial heat-transfer resistance and conjugate heat transfer to better reproduce experimental conditions \citep{burevs2022comprehensive,long2025direct}. In parallel, multiscale simulations have coupled micro-region models with macroscopic bubble-growth simulations to account for evaporation within the unresolved micro-region and its effects on heat transfer and the apparent contact angle \citep{torres2024coupling,wei2025effect}. Despite these advances, the near-contact-line region may remain insufficiently resolved in full nucleate-boiling DNS, as indicated by the step-like microlayer profiles and insufficiently smooth near-contact-line velocity fields reported in our previous study \citep{long2025direct}. These limitations make it difficult to distinguish the contributions of evaporation-induced flow and contact-line motion to the local liquid flow and interfacial-slope distribution.

A simplified dewetting-film configuration, which provides an idealized representation of the microlayer near a receding contact line, therefore offers a more controlled setting for investigating the coupled effects of evaporation and contact-line motion. Although lubrication models can retain both contributions over a broad range of length scales, they rely on simplified descriptions of the local flow and heat transfer. By resolving the two-dimensional flow and temperature fields together with the evolving interface, DNS provides an independent basis for examining the near-contact-line dynamics and assessing the scale-separated description without invoking the lubrication approximation.

In this paper, we perform two-dimensional direct numerical simulations of micrometre-scale film dewetting with phase change on a superheated solid wall. Finite interfacial heat-transfer resistance and Navier slip are incorporated to regularize the thermal and viscous singularities at the contact line. The physical model and numerical methods are described in \S\ref{sec:physcal_numerical}. In \S\ref{sec:3.1}, we first establish an isothermal-dewetting baseline by comparing the simulated contact-line speed, interfacial-slope distribution and apparent contact angle with the asymptotic theory of \citet{snoeijer2010asymptotic}. In \S\ref{sec:3.2}, we examine how phase change modifies the contact-line motion, interfacial heat flux, near-contact-line liquid flow and interfacial-slope distribution, assess the applicability of the scale-separated description, and derive a first-order evaporation correction to the Cox–Voinov solution. In \S\ref{sec:3.3}, the DNS results are compared with lubrication-model solutions retaining or neglecting contact-line motion. In \S\ref{sec:3.4} and \S\ref{sec:3.5}, we investigate the effects of equilibrium contact angle and slip length, respectively. Finally, \S\ref{sec:con} summarizes the main findings.

\section{Physical model and numerical methods}
\label{sec:physcal_numerical}

\subsection{Physical model}\label{sec:2.1}

%\begin{figure}[t]
%	\centerline{\includegraphics[width=\linewidth]{figs/fig2-1-1.pdf}}
%	%\captionsetup{singlelinecheck=off,justification=center}
%	\captionsetup{style=capcenter}
%	\caption{Schematic of a dewetting liquid film with phase change on a superheated solid wall. The $y$-axis is enlarged for clarity.}
%	\label{fig:2-1-1}
%\end{figure}

\begin{figure}[t]
	\centerline{\includegraphics[width=\linewidth]{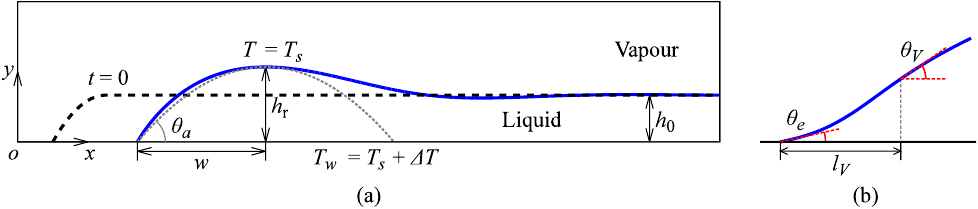}}
	%\captionsetup{singlelinecheck=off,justification=center}
%	\captionsetup{style=capcenter}
	\caption{Schematic of film dewetting with phase change on a superheated solid wall. (a) Film geometry with the $y$-axis enlarged for clarity. The black dashed line denotes the initial profile ($t=0$), while the grey dotted parabola, fitted using the rim height $h_r$ and half-width $w$, defines the apparent contact angle $\theta_a$. (b) Micro-region near the contact line, where the interfacial slope increases from the equilibrium contact angle $\theta_e$ to the Voinov angle $\theta_V$ over an effective length scale $l_V$.}
	\label{fig:2-1-1}
\end{figure}

We consider a two-dimensional static liquid film of initial thickness $h_0 = 5\,\mathrm{\mu m}$ deposited on a solid wall maintained at a superheat of $\Delta T$, as illustrated in figure~\ref{fig:2-1-1}. The film initially extends over a length of $475\,\mathrm{\mu m}$ within a computational domain of size $500\,\mathrm{\mu m} \times 500\,\mathrm{\mu m}$, and is surrounded by its own vapour. The initial liquid--vapour interface consists of a flat film terminated by a circular-arc rim that meets the solid wall at a prescribed equilibrium contact angle $\theta_e$. Driven by capillary forces, the contact line retracts spontaneously, leading to liquid accumulation near the film edge and the formation of a bulged rim, while phase change takes place simultaneously at the liquid--vapour interface. The liquid and vapour phases are taken to be saturated water and its vapour at $1\,\mathrm{atm}$, and their thermophysical properties are listed in table~\ref{tab:property}. Gravity is neglected because the film thickness is much smaller than the capillary length $l_c = [\sigma/(\rho_l g)]^{1/2}$, where $g$ is the gravitational acceleration. In this study, the equilibrium contact angle is varied in the range $5^\circ \leq \theta_e \leq 20^\circ$, while the wall superheat is varied  in the range $0 \leq \Delta T \leq 16\,\mathrm{K}$.

The dewetting dynamics of the liquid film with phase change are governed by the incompressible Navier--Stokes equations coupled with the temperature transport equation, namely
\begin{equation}\label{eq:nondiv}
	\nabla\cdot\boldsymbol{u} = S_{pc},
\end{equation}
\begin{equation}\label{eq:NS}
	\rho \left( \frac{\partial\boldsymbol{u}}{\partial t} + \boldsymbol{u}\cdot\nabla\boldsymbol{u} \right)
	= -\nabla p + \nabla\cdot(2\mu\mathbb{D}) + \boldsymbol{F}_s,
\end{equation}
\begin{equation}\label{eq:temp}
	\rho C_p \left( \frac{\partial T}{\partial t} + \boldsymbol{u} \cdot \nabla T \right)
	= \nabla \cdot \left( k \nabla T \right).
\end{equation}
Here, $\boldsymbol{u}$, $p$ and $T$ denote the velocity, pressure and temperature fields, respectively, while $\rho$, $\mu$, $k$ and $C_p$ are the density, dynamic viscosity, thermal conductivity and specific heat capacity of the fluid. The tensor $\mathbb{D}$ is the deformation-rate tensor, defined by $\mathbb{D}_{ij} = (\partial_j u_i + \partial_i u_j)/2$. The term $S_{pc}$ in \eqref{eq:nondiv} represents the volumetric source term induced by interfacial phase change. The surface tension force is written as $\boldsymbol{F}_s = -\sigma \kappa \delta_s \boldsymbol{n}$, where $\sigma$ is the surface tension coefficient, $\kappa$ is the interface curvature, $\delta_s$ is the surface Dirac delta function that localizes the force at the liquid--vapour interface, and $\boldsymbol{n}$ is the unit normal vector pointing from the liquid to the vapour phase. In the present study, $\sigma$ is assumed to be constant, and thus Marangoni effects are neglected. In addition, the heat generation due to viscous dissipation is neglected in the temperature equation \eqref{eq:temp}.

Across the liquid--vapour interface $\Gamma$, the discontinuities in pressure, velocity and heat flux are governed by the following interfacial jump conditions:
\begin{equation}\label{eq:jump-p}
	[p]_{\Gamma} = \sigma \kappa,
\end{equation}
\begin{equation}\label{eq:jump-u}
	[\boldsymbol{u}]_{\Gamma} = \dot{m} \left[ \frac{1}{\rho} \right]_{\Gamma} \boldsymbol{n},
\end{equation}
\begin{equation}\label{eq:jump-q}
	[q]_{\Gamma} = \dot{m} L_e.
\end{equation}
Here, $[\cdot]_{\Gamma}$ denotes the jump of any physical quantity $\phi$ across the interface, defined by $[\phi]_{\Gamma} = \phi_l - \phi_v$, where the subscripts $l$ and $v$ refer to the liquid and vapour phases, respectively. The quantity $\dot{m}$ is the interfacial mass flux, defined as
\begin{equation}\label{eq:mass-flux}
	\dot{m} = \rho_l \left( \boldsymbol{u}_l - \boldsymbol{u}_{\Gamma} \right) \cdot \boldsymbol{n}
	= \rho_v \left( \boldsymbol{u}_v - \boldsymbol{u}_{\Gamma} \right) \cdot \boldsymbol{n},
\end{equation}
where $\boldsymbol{u}_{\Gamma}$ is the interface velocity, which generally differs from the local liquid and vapour velocities in the presence of phase change. The heat flux $q$ in \eqref{eq:jump-q} for each phase is defined as $q = -k \nabla T \cdot \boldsymbol{n}$, with positive values directed from the liquid towards the vapour, and $L_e$ is the latent heat of phase change.

\begin{table}
  \begin{center}
	\def~{\hphantom{0}}
	  \begin{tabular}{lcc}
	  	&  Water                                                      &  Vapour      \\[3pt]
	  	Density $\rho$\;$\mathrm{(kg\,m^{-3})}$                    &  958          &  0.597 \\
	  	Dynamic viscosity $\mu$\;$\mathrm{(kg\,m^{-1}\,s^{-1})}$    & \quad\quad\quad\quad\quad\quad  $2.82 \times 10^{-4}$ \quad\quad\quad\quad\quad\quad  &  $ 1.22 \times 10^{-5}$ \\
	  	Thermal conductivity $k$\;$\mathrm{(W\,m^{-1}\,K^{-1})}$   &  $0.677$   &  $2.4 \times 10^{-2}$ \\
	  	Specific heat $C_p$\;$\mathrm{(J\,kg^{-1}\,K^{-1})}$   & 4216                &   2034 \\
	  	Surface tension $\sigma$\;$\mathrm{(N\,m^{-1})}$ & \multicolumn{2}{c}{\quad\quad\quad\quad\quad\quad  $5.89 \times 10^{-2}$} \\
	  	Latent heat $L_e$\;$\mathrm{(J\,kg^{-1})}$ & \multicolumn{2}{c}{\quad\quad\quad\quad\quad\quad  $2.26 \times 10^{6}$} \\
	  	Saturation temperature $T_s$\;$\mathrm{(K)}$ & \multicolumn{2}{c}{\quad\quad\quad\quad\quad\quad  $373$} \\
	  	Wall superheat $\mathrm{\Delta} T$\;$\mathrm{(K)}$ & \multicolumn{2}{c}{\quad\quad\quad\quad\quad\quad  $0\sim 16$} \\
	  	Equilibrium contact angle $\theta_e$\;$\mathrm{(^\circ)}$  & \multicolumn{2}{c}{\quad\quad\quad\quad\quad\quad  $5\sim 20$} \\
	  	Accommodation coefficient $f_{ac}$  & \multicolumn{2}{c}{\quad\quad\quad\quad\quad\quad  $0.05\sim 0.4$} 
	  \end{tabular}
	  \caption{Thermophysical properties of water and vapour, and parameter ranges considered in the simulations.}
	  \label{tab:property}
	  \end{center}
\end{table}

Equation \eqref{eq:jump-p} represents the capillary pressure jump across the liquid--vapour interface, neglecting corrections associated with the viscous normal-stress jump and phase-change-induced momentum flux. In the one-fluid formulation, the pressure jump is incorporated into the momentum equation \eqref{eq:NS} through the surface-tension force $\boldsymbol{F}_s$. The velocity jump \eqref{eq:jump-u} follows from the interfacial mass balance \eqref{eq:mass-flux}, and gives rise to the volumetric source term appearing on the right-hand side of the continuity equation \eqref{eq:nondiv}, namely
\begin{equation}
	S_{pc} = -\dot{m} \left[\frac{1}{\rho}\right]_{\Gamma}\delta_s.
\end{equation}
The heat-flux jump condition \eqref{eq:jump-q} is obtained from the interfacial energy balance. In the absence of additional kinetic resistance \citep{janevcek2013moving}, the local thermodynamic equilibrium at the liquid--vapour interface would imply
\begin{equation}\label{eq:temperature}
	T_{l,\Gamma} = T_{v,\Gamma} = T_s,
\end{equation}
where $T_{l,\Gamma}$ and $T_{v,\Gamma}$ denote the temperatures on the liquid and vapour sides of the interface, respectively, and $T_s$ is the saturation temperature of the liquid--vapour system at the ambient pressure. Combining \eqref{eq:jump-q} and \eqref{eq:temperature}, the interfacial mass flux $\dot{m}$ can be determined by the heat fluxes reaching the interface from both liquid and vapour sides.

In addition to the interfacial jump conditions described above, special treatment is also required in the vicinity of the contact line, where two distinct singularities arise in dewetting with phase change. The first is associated with the contact line moving along a no-slip solid wall, which leads to a non-integrable viscous stress near the contact line \citep{huh1971hydrodynamic}. To regularize this singularity, a Navier slip boundary condition is imposed for the tangential velocity at the solid wall. In the coordinate system shown in figure~\ref{fig:2-1-1}, this condition is written as
\begin{equation}\label{eq:navier}
	u_x - U_w = \lambda_s \frac{\partial u_x}{\partial y}.
\end{equation}
Here, $U_w$ is the wall velocity, which is set to zero in the present study, and $\lambda_s$ is the slip length. Unless otherwise specified, $\lambda_s$ is fixed at $500\,\mathrm{nm}$. This value is significantly larger than the physical slip length, which is typically of molecular scale \citep{snoeijer2013moving}. Such a choice is dictated by the grid resolution affordable in the present simulations, since the grid spacing must be smaller than the prescribed slip length in order to resolve the slip region \citep{afkhami2009mesh}.

The second singularity concerns the temperature field. With the liquid--vapour interface maintained at the saturation temperature $T_s$ and the solid wall held at a constant superheat of $\Delta T$, the conductive heat flux through the liquid film scales as $q \sim k_l \Delta T/h$ and therefore diverges as the local film thickness $h$ tends to zero near the contact line. To regularize this thermal singularity, we account for a finite interfacial heat transfer resistance (IHTR), which represents the finite kinetic resistance associated with evaporation and condensation at the liquid surface. At the constitutive level, the interfacial heat flux may be related to the temperature difference across the interface by \citep{giustini2016evaporative}
\begin{equation}\label{eq:flux}
	q_l = \frac{T_{l,\Gamma} - T_{v,\Gamma}}{R_{\Gamma}}
	= \frac{T_{l,\Gamma} - T_s}{R_{\Gamma}},
\end{equation}
where an assumption of $T_{v,\Gamma} = T_s$ is adopted. With a finite interfacial thermal resistance, a non-zero heat flux gives rise to a temperature discontinuity across the liquid--vapour interface. Therefore, the thermodynamic equilibrium condition \eqref{eq:temperature} no longer applies. The interfacial thermal resistance $R_{\Gamma}$ is given by \citep{marengo2021surface}
\begin{equation}
	R_{\Gamma} = \frac{2-f_{ac}}{2f_{ac}} \frac{T_s}{\rho_v L_e^2}\sqrt{\frac{2\pi R_g T_s}{M}},
\end{equation}
where $R_g$ is the universal gas constant and $M$ is the molar mass of water. Here, $f_{ac}$ is the accommodation coefficient accounting for the scattering processes associated with both molecular emission from and absorption into the liquid surface \citep{giustini2016evaporative}. The coefficient $f_{ac}$ takes values between 0 and 1. However, despite its central role in determining $R_{\Gamma}$, no consensus has yet been reached on its value for the water--vapour system, and the values inferred from experimental data vary substantially among different studies \citep{tecchio2024microlayer}. Thus, in the present study, the accommodation coefficient is varied in the range $0.05 \leq f_{ac} \leq 0.4$.

Since heat transfer within the dewetting film is dominated by the conduction in the wall-normal direction, the temperature field in the liquid film can be approximated by one-dimensional steady heat conduction. By balancing the interfacial heat flux \eqref{eq:flux} with the conductive heat flux through a liquid film of thickness $h$, we obtain
\begin{equation}\label{eq:flux1}
	q_l 
	= \frac{T_w - T_{l,\Gamma}}{h/k_l}
	= \frac{T_{l,\Gamma} - T_s}{R_{\Gamma}}
	\equiv \frac{T_w - T_s}{h/k_l + R_{\Gamma}},
\end{equation}
where $T_w = T_s + \Delta T$ is the wall temperature. Equation \eqref{eq:flux1} shows that the finite IHTR removes the divergence of heat flux as $h \to 0$, yielding a bounded heat flux $(T_w-T_s)/R_\Gamma$ at the contact line. 

To implement the IHTR numerically, however, we do not explicitly resolve the temperature jump on the liquid side of the interface. Instead, following \citet{burevs2022comprehensive} and \citet{MAGNINI2026128926}, the interfacial thermal resistance is represented at the fluid--solid boundary by an equivalent contact heat-transfer resistance equal to $R_{\Gamma}$. The heat-flux balance is then written as
\begin{equation}\label{eq:flux2}
	q_l = \frac{T_{l,w} - T_s}{h/k_l}
	= \frac{T_w - T_{l,w}}{R_{\Gamma}}
	\equiv \frac{T_w - T_s}{h/k_l + R_{\Gamma}},
\end{equation}
where $T_{l,w}$ is the liquid temperature at the solid wall, which differs from $T_w$ due to the thermal resistance $R_{\Gamma}$. Comparing \eqref{eq:flux1} and \eqref{eq:flux2}, one finds that both formulations yield the same liquid-side heat flux. The equivalent wall-resistance formulation therefore retains the regularizing effect of the interfacial thermal resistance while allowing the equilibrium condition \eqref{eq:temperature} to be imposed at the liquid--vapour interface, thereby simplifying the numerical implementation.

\begin{figure}[t]
	\centering
	\includegraphics[width=0.5\linewidth]{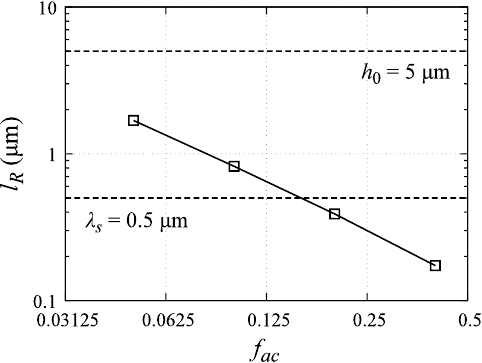}
	\caption{Variation of the thermal-resistance length scale $l_R = k_l R_{\Gamma}$ with the accommodation coefficient $f_{ac}$. The initial film thickness $h_0$ and the slip length $\lambda_s$ are also shown for comparison.}
	\label{fig:2-1-2}
\end{figure}

From \eqref{eq:flux2}, a length scale associated with the interfacial heat-transfer resistance can be defined as
\begin{equation}\label{eq:thermal-length}
	l_R = k_l R_{\Gamma}.
\end{equation}
This length may be interpreted as the thickness of an equivalent liquid layer having the same thermal resistance as the IHTR. The variation of $l_R$ with the accommodation coefficient $f_{ac}$ over the range considered in the present study is shown in figure~\ref{fig:2-1-2}. As expected, $l_R$ decreases monotonically with increasing $f_{ac}$. For comparison, two other characteristic length scales relevant to the present problem, namely the initial film thickness $h_0$ and the slip length $\lambda_s$, are also shown in figure~\ref{fig:2-1-2}. To obtain grid-independent results, the numerical mesh must be sufficiently fine to resolve these three characteristic lengths, in particular the smaller of $l_R$ and $\lambda_s$.

For completeness, we also assess two effects that are neglected in the present model, namely the Kelvin effect and vapour recoil. The Kelvin effect accounts for the dependence of the saturation temperature on interface curvature \citep{rednikov2013singularity}. Its associated temperature shift can be estimated as $\Delta T_K = {\sigma \kappa T_s}/{\rho_l L_e}$. Since the interface curvature remains well below $1/\Delta$ in the simulations, where $\Delta$ denotes the grid size, it follows that $\Delta T_K \ll {\sigma T_s}/{\rho_l L_e \Delta} \approx 0.1\,\mathrm{K}$. This temperature shift remains much smaller than the wall superheat considered here. The Kelvin effect is thus neglected. Another effect omitted here is vapour recoil, which gives rise to an additional normal stress at the interface \citep{nikolayev2010dynamics}. Its magnitude can be estimated as $\Delta p_r = \dot{m}^2 \left( {1}/{\rho_v} - {1}/{\rho_l} \right)$. Using $\dot{m} \sim q_l/L_e$ together with \eqref{eq:flux2}, and taking $h \to 0$ to obtain an upper-bound estimate of the heat flux, the ratio of vapour recoil pressure to capillary pressure is estimated as
\begin{equation}
	\frac{\Delta p_r}{\sigma \kappa_c}
	\ll
	\frac{\Delta T^2 h_0}{L_e^2 R_{\Gamma}^2 \rho_v \sigma}
	\approx 0.1,
\end{equation}
for the largest wall superheat and the smallest $R_\Gamma$ considered in the present simulations, where $\kappa_c \sim 1/h_0$ is used as an estimate of the characteristic curvature. Vapour recoil is therefore expected to play only a minor role and is neglected in the present model.

To complete the specification of the present physical model, we summarize the boundary and initial conditions used in the simulations. As shown in figure~\ref{fig:2-1-1}, for the velocity field, outflow boundary conditions are imposed at the top and right boundaries, while symmetry boundary conditions are imposed at the left boundary. At the bottom boundary, the tangential velocity satisfies the Navier-slip condition \eqref{eq:navier}, while the normal velocity vanishes at the solid wall. For the temperature field, adiabatic boundary conditions are imposed at the left and right boundaries, while the temperature at the top boundary is fixed at $T=T_s$. At the bottom boundary, the wall temperature is prescribed as $T_w=T_s+\Delta T$. For the liquid phase, the equivalent thermal resistance is imposed through the following Robin boundary condition:
\begin{equation}\label{eq:robin}
	-k_l\left.\frac{\partial T_l}{\partial y}\right|_w
	=\frac{T_w-T_{l,w}}{R_\Gamma}, 
\end{equation}
whereas, for the vapour phase, a Dirichlet boundary condition of $T_v = T_w$ is applied. In addition, a constant equilibrium contact angle $\theta_e$ is prescribed at the solid wall and is taken as the microscopic contact angle in the present model. Initially, both the liquid and vapour phases are at rest, and the temperature field is initialized uniformly at the saturation temperature $T_s$.

\subsection{Numerical methods}

The governing equations, \eqref{eq:nondiv}--\eqref{eq:temp}, are solved using the open-source finite-volume solver \textit{Basilisk} \citep{popinet2009accurate,popinet2015quadtree}, which employs a collocated variable arrangement and a second-order approximate projection method for the pressure--velocity coupling \citep{bell1989second}. The liquid--vapour interface is captured by a conservative volume-of-fluid (VOF) method \citep{weymouth2010conservative}, in which the interface is reconstructed using the piecewise linear interface construction (PLIC) technique and advected geometrically. Surface tension is incorporated through the continuum surface force (CSF) formulation \citep{brackbill1992continuum}. Combined with the balanced-force implementation and the height-function-based curvature evaluation \citep{cummins2005estimating}, this treatment effectively reduces parasitic currents near the interface \citep{popinet2018numerical}. In addition, adaptive mesh refinement (AMR) is employed to concentrate the resolution where needed. In the present computations, cells in the vicinity of the liquid--vapour interface and, in particular, near the moving contact line are selectively refined according to the gradients of the liquid volume fraction and velocity fields. This strategy allows the thin film, the rim region and the contact-line region to be resolved accurately while keeping the overall computational cost affordable. Further details on the application and validation of \textit{Basilisk} for interfacial flows, including cases with phase change, can be found in \citet{vleeschhouwers2026liquid,long2025direct,maes2025birth,kulkarni2025atomising,zhao2025direct}.

In the continuity equation \eqref{eq:nondiv}, phase change introduces a mass source term that gives rise to the velocity jump described by \eqref{eq:jump-u}. Since this source term is singular and confined to interfacial cells, the classical approximate projection method may produce oscillations in the velocity field \citep{zhao2022boiling,long2024edge}. To suppress these oscillations, the ghost fluid method is employed \citep{tanguy2014benchmarks}. For the temperature transport equation \eqref{eq:temp}, an operator-splitting approach is adopted \citep{cipriano2024multicomponent}. In the first step, the temperature advection is performed using a geometric flux evaluation consistent with the VOF advection procedure. In the second step, thermal diffusion is solved separately in each phase, with additional source terms incorporated in interfacial cells to account for phase change. This treatment takes into account the jump in the temperature gradient across the interface, as expressed by \eqref{eq:jump-q}, and effectively limits numerical diffusion of temperature. Further details of the numerical treatment of phase change can be found in \citet{cipriano2024multicomponent,long2025direct}. In addition, because very small contact angles are considered in the present study, a newly developed height-function-based contact-angle model is adopted \citep{han2026applying}. Several additional numerical treatments specific to the present dewetting-film simulations with phase change are described below.

\subsubsection{Boundary treatment for the level-set function}

In the ghost fluid method (GFM) \citep{tanguy2007level,tanguy2014benchmarks}, two ghost velocity fields are introduced for the liquid and vapour phases based on the velocity jump condition \eqref{eq:jump-u}, namely
\begin{subequations}\label{eq:gfm}
	\begin{equation}
		\boldsymbol{u}_l^g = \boldsymbol{u}_v + \dot{m} \left[\frac{1}{\rho}\right]_{\Gamma}\boldsymbol{n}
		\qquad \text{for} \quad \phi > 0,
	\end{equation}
	\begin{equation}
		\boldsymbol{u}_v^g = \boldsymbol{u}_l - \dot{m} \left[\frac{1}{\rho}\right]_{\Gamma}\boldsymbol{n}
		\qquad \text{for} \quad \phi < 0.
	\end{equation}
\end{subequations}
Here, the superscript $g$ denotes a ghost quantity. The level-set function $\phi$ is constructed as the signed distance from the cell centre to the interface reconstructed from the VOF field \citep{LIMARE2023111829}, and is defined to be positive in the vapour phase. Using \eqref{eq:gfm}, two auxiliary velocity fields can then be defined over the whole computational domain as
\begin{align}\label{eq:gfm2}
	\tilde{\boldsymbol{u}}_l =
	\begin{cases}
		\boldsymbol{u}_l, & \phi<0,\\
		\boldsymbol{u}_l^g, & \phi>0,
	\end{cases}
	\qquad
	\tilde{\boldsymbol{u}}_v =
	\begin{cases}
		\boldsymbol{u}_v, & \phi>0,\\
		\boldsymbol{u}_v^g, & \phi<0.
	\end{cases}
\end{align}
With the ghost velocities defined by \eqref{eq:gfm}, the resulting auxiliary velocity fields $\tilde{\boldsymbol{u}}_l$ and $\tilde{\boldsymbol{u}}_v$ are continuous across the interface, thereby removing the singularity associated with the interfacial mass source term. The evaluation of $\boldsymbol{u}_l^g$ and $\boldsymbol{u}_v^g$ requires the interface normal $\boldsymbol{n}$ to be defined throughout the computational domain. This normal vector can be obtained from the gradient of the level-set function $\nabla\phi$. Moreover, the interfacial mass flux $\dot{m}$ must be extrapolated away from the interface into both phases using PDE-based extrapolation techniques \citep{aslam2004partial}, which also relies on the gradient of $\phi$. The accurate construction of the level-set function is therefore essential to the present implementation.

\begin{figure}[t]
	\centering
	\begin{subfigure}[t]{0.48\textwidth}
		\centering
		\includegraphics[width=\linewidth]{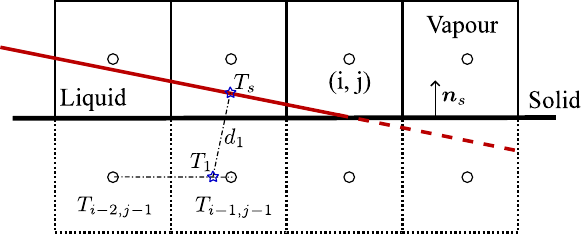}
		\subcaption{}
	\end{subfigure}
	\hfill
	\begin{subfigure}[t]{0.49\textwidth}
		\centering
		\includegraphics[width=\linewidth]{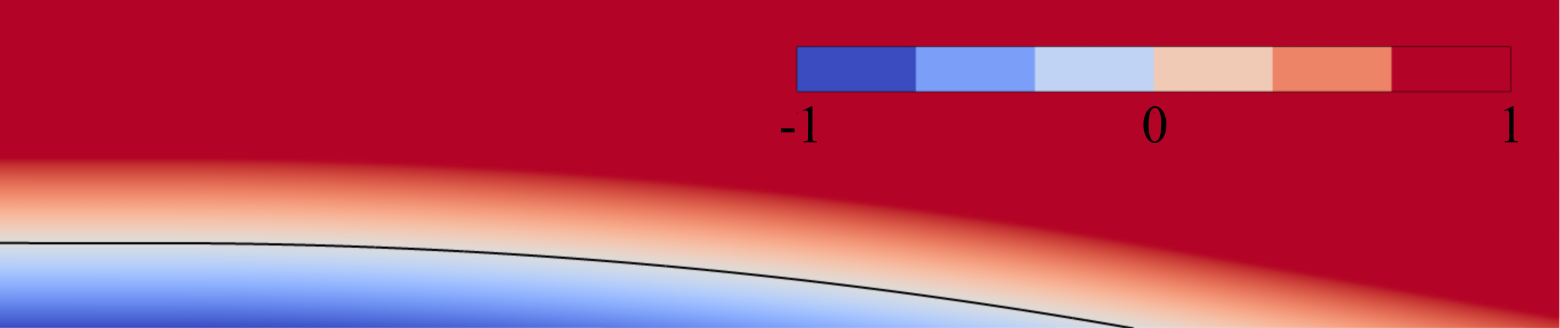}
		\subcaption{}
	\end{subfigure}
	\begin{subfigure}[t]{0.49\textwidth}
		\centering
		\includegraphics[width=\linewidth]{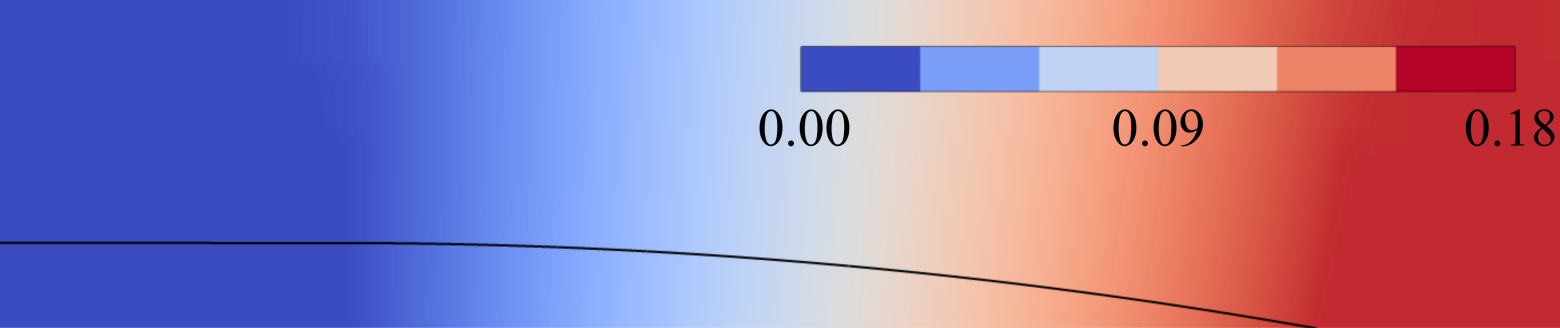}
		\subcaption{}
	\end{subfigure}
	\hfill
	\begin{subfigure}[t]{0.49\textwidth}
		\centering
		\includegraphics[width=\linewidth]{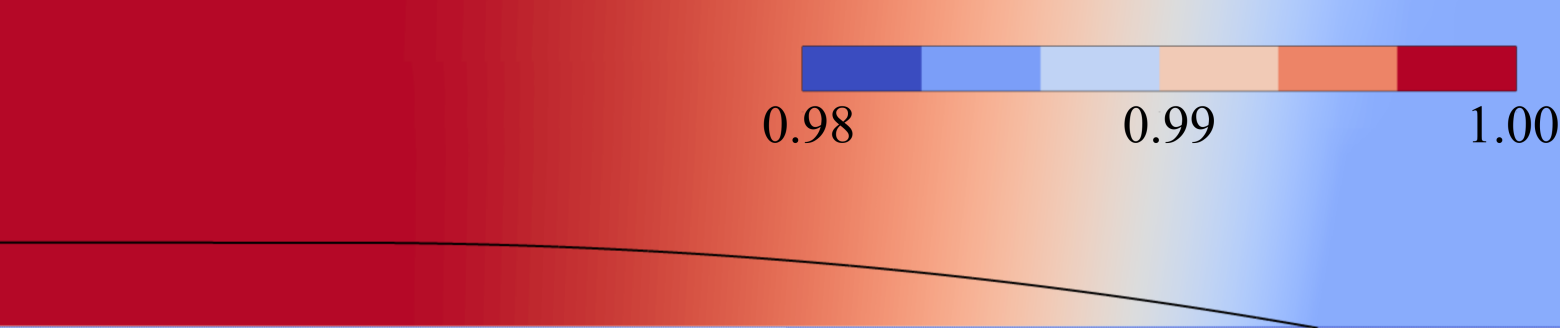}
		\subcaption{}
	\end{subfigure}
	\caption{Boundary treatment for the level-set function at $\theta_e = 10^\circ$. (a) Schematic illustration of the boundary treatment near the contact line. (b) Contours of the normalised level-set function, $\phi/h_0$. (c) Contours of $\partial \phi/\partial x$. (d) Contours of $\partial \phi/\partial y$.}
	\label{fig:2-2-1}
\end{figure}

Special care is required near the solid wall because the present dewetting simulations involve very small contact angles. As illustrated in figure~\ref{fig:2-2-1}(a), the interface in the contact-line cell $(i,j)$ is extended into the solid by imposing
\begin{equation}
	\boldsymbol{n}_s \cdot \nabla \phi = \cos \theta_e,
\end{equation}
where $\boldsymbol{n}_s$ is the unit normal vector to the solid surface. With this extended interface, the level-set values in the cells around the contact line can be well determined, as shown in figure~\ref{fig:2-2-1}(b). In addition, values of the level-set function must also be prescribed in the ghost cells (the dashed cells in figure~\ref{fig:2-2-1}(a)), since they are required to evaluate $\nabla \phi$ in those cells adjacent to the solid boundary. The resulting fields $\partial \phi/\partial x$ and $\partial \phi/\partial y$ are shown in figure~\ref{fig:2-2-1}(c,d), and vary smoothly near the contact line.

\subsubsection{Liquid-side interfacial temperature gradient near the contact line}

The evaluation of the interfacial mass flux $\dot{m}$ requires the heat fluxes on both sides of the interface, as indicated by \eqref{eq:jump-q}. This jump condition also implies that the temperature gradient is discontinuous across the interface. Therefore, the heat flux in each phase must be evaluated using only the temperature information from the corresponding side of the interface \citep{cipriano2024multicomponent}. In the present dewetting-film simulations, special treatment is required to evaluate the liquid-side temperature gradient near the contact line because of the very small contact angles involved.

As illustrated in figure~\ref{fig:2-2-1}(a), the interfacial cell $(i-1,j)$ is taken as an example. To evaluate the liquid-side temperature gradient, we start from the midpoint of the reconstructed interface, where the temperature is equal to the saturation temperature $T_s$, and then move along the interface normal into the liquid by a distance $d_1$ to locate the point at which the temperature $T_1$ is to be evaluated. The temperature $T_1$ at this point is then obtained by linear interpolation from the two neighbouring ghost-cell temperatures, $T_{i-2,j-1}$ and $T_{i-1,j-1}$. These ghost-cell temperatures are determined by discretizing the temperature boundary condition \eqref{eq:robin}. For example,
\begin{equation}
	T_{i-1,j-1}
	= 2T_{l,w} - T_{i-1,j}
	= 2\,\frac{T_w \Delta + 2T_{i-1,j} k_l R_{\Gamma}}{\Delta + 2k_l R_{\Gamma}} - T_{i-1,j}.
\end{equation}
The liquid-side heat flux is then estimated as
\begin{equation}\label{eq:one-side}
	q_l = k_l \frac{T_1 - T_s}{d_1}.
\end{equation}
This evaluation is only first-order accurate. It may be regarded as a degenerate case of the heat-flux evaluation used away from the solid wall, where a second-order estimate can be applied \citep{cipriano2024multicomponent}.

\subsubsection{A modification to the ghost fluid method}
\label{sec:gfm}

\begin{figure}[t]
	\centering
	\begin{subfigure}[t]{0.48\textwidth}
		\centering
		\includegraphics[width=\linewidth]{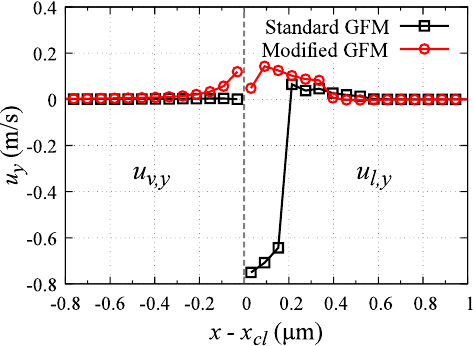}
		\subcaption{}
	\end{subfigure}
	\hfill
	\begin{subfigure}[t]{0.48\textwidth}
		\centering
		\includegraphics[width=\linewidth]{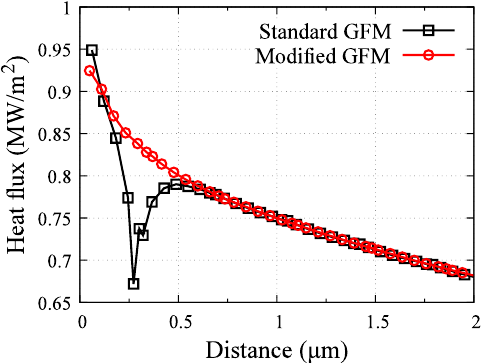}
		\subcaption{}
	\end{subfigure}
	\caption{Comparison between the standard and modified ghost fluid methods at $\theta_e=10^{\circ}$, $f_{ac}=0.05$ and $\Delta T = 2\,\mathrm{K}$. (a) Distribution of $u_y$ in the cells adjacent to the solid wall as a function of $x-x_{cl}$, where $x_{cl}$ denotes the contact-line position. Here, $u_{v,y}$ is shown for $x<x_{cl}$, whereas $u_{l,y}$ is shown for $x>x_{cl}$. The dashed line indicates the contact-line position. (b) Interfacial heat-flux distribution as a function of the horizontal distance $x - x_{cl}$ from the contact line.}
	\label{fig:2-2-2}
\end{figure}

In the standard ghost fluid method, the phase to which a grid cell belongs is determined by the sign of the level-set function, as indicated by \eqref{eq:gfm2}. The velocity jump condition \eqref{eq:gfm} is then used to reconstruct the ghost velocity of the other phase in both liquid and vapour cells. However, this procedure becomes problematic in the vicinity of the contact line when the contact angle is small. As illustrated in figure~\ref{fig:2-2-1}(a), the three interfacial cells closest to the contact line, namely $(i,j)$, $(i-1,j)$ and $(i-2,j)$, are all identified as vapour cells since $\phi>0$ in these cells. If the wall-normal vapour velocity $u_{v,y}$ satisfies a no-penetration condition at the solid wall, the jump condition \eqref{eq:gfm} then yields a negative ghost liquid normal velocity in these cells. For a representative case with $\theta_e = 10^{\circ}$, the resulting near-wall distribution of liquid normal velocity is shown in figure~\ref{fig:2-2-2}(a), together with the corresponding vapour normal velocity. It can be seen that, on the liquid side of the contact line ($x>x_{cl}$), the liquid normal velocity attains a large negative value in the immediate vicinity of the contact line. This behaviour arises from the ghost liquid velocity reconstruction and is clearly non-physical. Such a spurious near-wall liquid velocity contaminates the interface advection and, in turn, produces a non-monotonic heat-flux distribution around the contact line, as shown in figure~\ref{fig:2-2-2}(b).

%\begin{figure}[t]
%	\centering
%	\begin{subfigure}[t]{0.48\textwidth}
%		\centering
%		\includegraphics[width=\linewidth]{figs/fig2-2-3a.pdf}
%		\subcaption{}
%	\end{subfigure}
%	\hfill
%	\begin{subfigure}[t]{0.48\textwidth}
%		\centering
%		\includegraphics[width=\linewidth]{figs/fig2-2-3b.pdf}
%		\subcaption{}
%	\end{subfigure}
%	\caption{
%	Influence of the vapour-phase velocity boundary condition on the near-wall velocity distribution and interface slope at $\theta_e=10^{\circ}$, $f_{ac}=0.05$ and $\Delta T = 2\,\mathrm{K}$. (a) Distribution of $u_y$ in the cells adjacent to the solid wall as a function of $x-x_{cl}$, where $x_{cl}$ denotes the contact-line position. The legend labels `no-penetration' and `outflow' denote the boundary conditions imposed on the vapour normal velocity $u_{v,y}$ at the wall. The region between the two dashed lines corresponds to interfacial cells with the volume fraction $f$ satisfying $0 < f < 1$. (b) Interfacial slope distribution as a function of the distance from the contact line.}
%	\label{fig:2-2-3}
%\end{figure}

To eliminate this spurious liquid normal velocity near the contact line, we modify the phase-classification criterion used in the ghost-velocity reconstruction. Specifically, cells are classified as liquid or vapour according to the volume fraction $f$, rather than according to the sign of the level-set function $\phi$. With this criterion, the ghost velocities are constructed as
\begin{subequations}\label{eq:gfm-new}
	\begin{equation}
		\boldsymbol{u}_l^g = \boldsymbol{u}_v + \dot{m} \left[\frac{1}{\rho}\right]_{\Gamma}\boldsymbol{n}
		\qquad \text{for} \quad f = 0,
	\end{equation}
	\begin{equation}
		\boldsymbol{u}_v^g = \boldsymbol{u}_l - \dot{m} \left[\frac{1}{\rho}\right]_{\Gamma}\boldsymbol{n}
		\qquad \text{for} \quad f > 0.
	\end{equation}
\end{subequations}
The same volume-fraction-based criterion is also used when constructing the auxiliary velocity fields in \eqref{eq:gfm2}. Compared with the standard ghost fluid method, in which the phase classification of grid cells is based on the sign of the level-set function and therefore follows the reconstructed interface, the modified treatment classifies all cells with non-zero liquid volume fraction as liquid cells for the purpose of the GFM reconstruction. This can be interpreted as shifting the cell-classification boundary by approximately one grid spacing $\Delta$ into the vapour phase. Reconsidering the configuration shown in figure~\ref{fig:2-2-1}(a), the three interfacial cells closest to the contact line, namely $(i,j)$, $(i-1,j)$ and $(i-2,j)$, are therefore classified as liquid cells because $f>0$ in these cells. The resulting near-wall distributions of the liquid and vapour normal velocities are shown in figure~\ref{fig:2-2-2}(a). The spurious negative liquid normal velocity obtained with the standard treatment is removed, and the corresponding interfacial heat-flux distribution becomes monotonic, as shown in figure~\ref{fig:2-2-2}(b).

\subsection{Grid independence study}

%\begin{figure}[t]
%	\centering
%	\begin{subfigure}[t]{0.48\textwidth}
%		\centering
%		\includegraphics[width=\linewidth]{figs/fig2-3-3a.pdf}
%		\subcaption{}
%	\end{subfigure}
%	\hfill
%	\begin{subfigure}[t]{0.48\textwidth}
%		\centering
%		\includegraphics[width=\linewidth]{figs/fig2-3-3b.pdf}
%		\subcaption{}
%	\end{subfigure}
%	\begin{subfigure}[t]{0.48\textwidth}
%		\centering
%		\includegraphics[width=\linewidth]{figs/fig2-3-3c.pdf}
%		\subcaption{}
%	\end{subfigure}
%	\hfill
%	\begin{subfigure}[t]{0.48\textwidth}
%		\centering
%		\includegraphics[width=\linewidth]{figs/fig2-3-3d.pdf}
%		\subcaption{}
%	\end{subfigure}
%	\caption{Grid-convergence study of the interfacial heat flux (a,b) and interface slope (c,d) versus the distance from the contact line at $t = 0.4\,\mathrm{ms}$ and $\Delta T = 8\,\mathrm{K}$. (a,c) $\theta_e = 10^{\circ}$ and $f_{ac} = 0.1$. (b,d) $\theta_e = 5^{\circ}$ and $f_{ac} = 0.4$.}
%	\label{fig:2-3-3}
%\end{figure}

In this section, we assess the grid independence of the simulations of film dewetting with phase change. The convergence study is carried out for the representative case with $\theta_e = 5^{\circ}$, $f_{ac} = 0.4$ and $\Delta T = 8\,\mathrm{K}$. Among the parameter combinations considered in the present study, this case corresponds to the smallest equilibrium contact angle and the smallest thermal-resistance length scale $l_R$ (see figure~\ref{fig:2-1-2}). Since $l_R$ decreases with increasing $f_{ac}$, the case with $f_{ac}=0.4$ imposes a stricter resolution requirement than cases with smaller accommodation coefficients. This choice of parameters gives rise to a pronounced influence of phase change on the near-contact-line dynamics, and therefore provides a demanding test of grid convergence. The grid resolution is varied over a wide range, from 5 to 80 grid cells across the initial film thickness $h_0$.

\begin{figure}[t]
	\centering
	\begin{subfigure}[t]{0.48\textwidth}
		\centering
		\includegraphics[width=\linewidth]{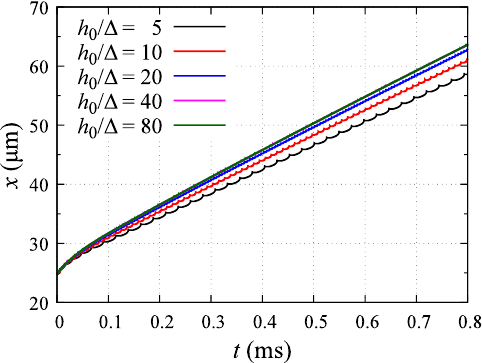}
		\subcaption{}
	\end{subfigure}
	\hfill
	\begin{subfigure}[t]{0.48\textwidth}
		\centering
		\includegraphics[width=\linewidth]{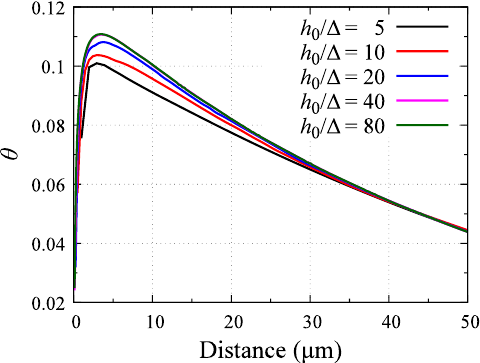}
		\subcaption{}
	\end{subfigure}
	\caption{Grid-convergence study for $\theta_e = 5^{\circ}$, $f_{ac} = 0.4$ and $\Delta T = 8\,\mathrm{K}$. Results are shown for different grid resolutions. (a) Time evolution of the contact-line position. (b) Interfacial slope as a function of the distance from the contact-line position at $t = 0.4\,\mathrm{ms}$.}
	\label{fig:2-3-1}
\end{figure}

Figure~\ref{fig:2-3-1}(a) shows the time evolution of the contact-line position for the grid resolutions considered. As the grid is refined, the numerical results converge towards the finest-grid solution. In particular, the curves obtained with $h_0/\Delta = 40$ and $h_0/\Delta = 80$ nearly overlap. For the present problem, grid convergence requires the resolution of the two microscopic length scales entering the model, namely the slip length $\lambda_s$ and the thermal-resistance length scale $l_R$. In the present case with $f_{ac}=0.4$, $l_R$ is the more restrictive scale; even on the finest grid, it is resolved by only about three grid cells. The close agreement between the $h_0/\Delta = 40$ and $h_0/\Delta = 80$ results therefore indicates that the contact-line motion is sufficiently resolved at $h_0/\Delta = 40$ for the purposes of the present study.

%Figure~\ref{fig:2-3-2} shows the film profiles at $t = 0.4\,\mathrm{ms}$ for the two representative cases. Consistent with the results shown in figure~\ref{fig:2-3-1}, the convergence in case (a) is better than that in case (b), and the profiles obtained with the two finest resolutions nearly overlap in both cases. A closer inspection of the interfacial shape near the contact line in case (b) shows that unphysical interfacial breakup occurs at coarse resolutions with $h_0/\Delta < 20$. This behaviour is likely caused by the large and highly localised heat flux in the vicinity of the contact line. At these resolutions, fewer than one grid cell spans the thermal-resistance length scale $l_R$ in case (b). This again highlights the importance of resolving both microscopic length scales, $\lambda_s$ and $l_R$, to obtain accurate and grid-converged numerical results.

Figure~\ref{fig:2-3-1}(b) shows the interfacial-slope distributions at $t = 0.4\,\mathrm{ms}$, which are plotted against the horizontal distance $x-x_{cl}$ from the instantaneous contact line. Hereafter, the local interfacial slope is defined as $\theta = \partial h/\partial x$, which approximates the local interface inclination angle under the small-slope assumption. In the DNS, this derivative is evaluated from the extracted interface profile using a centred finite difference. Compared with the contact-line position, the interfacial slope is a more local quantity and is therefore more sensitive to grid refinement. The main grid sensitivity is confined to the near-contact-line region: with increasing resolution, the maximum interfacial slope increases gradually, whereas the slope within the first few cells adjacent to the contact line decreases more noticeably. This behaviour is consistent with the fact that finer grids resolve smaller film thicknesses near the contact line. According to the local heat-flux relation \eqref{eq:flux2}, a smaller film thickness leads to a larger local heat flux and hence to a stronger phase-change-induced modification of the interfacial slope. This mechanism will be discussed in more detail in \S\ref{sec:3.2}. The slope distributions obtained with the two finest resolutions, $h_0/\Delta = 40$ and $h_0/\Delta = 80$, are nearly indistinguishable, with only a minor difference remaining very close to the contact line. Since the maximum interfacial slope is closely related to the capillary driving force and hence to the dewetting speed, the above convergence behaviour indicates that a resolution of $h_0/\Delta = 40$ is sufficient for the present simulations. This resolution is therefore adopted throughout the paper unless otherwise stated.

%For the case $\theta_e = 10^{\circ}$ and $f_{ac} = 0.1$, both quantities exhibit good grid convergence even for relatively coarse grids with $h_0/\Delta < 10$, as shown in figure~\ref{fig:2-3-3}(a,c). It should nevertheless be noted that the heat flux in the immediate vicinity of the contact line remains sensitive to the grid resolution. First, the heat flux in the contact-line cell depends strongly on the position of the geometrically reconstructed interface, as illustrated in figure~\ref{fig:2-2-1}(a). Secondly, as the grid is refined, the cell size decreases, leading to a smaller liquid-film thickness in the contact-line cell and hence to a larger local heat flux. For the case $\theta_e = 5^{\circ}$ and $f_{ac} = 0.4$, both the heat-flux and interface-slope distributions become irregular near the contact line on grids coarser than $h_0/\Delta = 20$, as shown in figure~\ref{fig:2-3-3}(b,d). This behaviour is consistent with the film profiles shown in figure~\ref{fig:2-3-2}(b). By contrast, at the two finest resolutions, the corresponding curves overlap closely. 

\section{Results}\label{Results}

\begin{figure}[t]
	\centering
	\begin{subfigure}[t]{\textwidth}
		\centering
		\includegraphics[width=\linewidth]{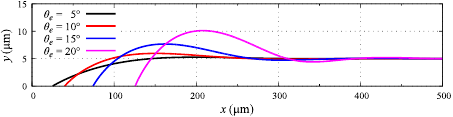}
		\subcaption{}
	\end{subfigure}
	\begin{subfigure}[t]{\textwidth}
		\centering
		\includegraphics[width=\linewidth]{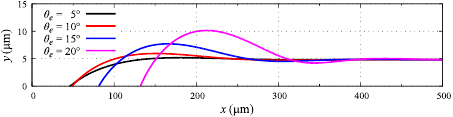}
		\subcaption{}
	\end{subfigure}
	\caption{Film profiles at $t = 0.4\,\mathrm{ms}$ for cases without (a) and with (b) phase change; the $y$ axis is enlarged for clarity. In panel (b), $\Delta T = 8\,\mathrm{K}$ and $f_{ac} = 0.4$.}
	\label{fig:3-1-1}
\end{figure}

In this section, we present the direct numerical simulation results for film dewetting with and without phase change. We first consider the isothermal case and compare the results with the asymptotic theory of \citet{snoeijer2010asymptotic}, which provides both a validation of the present numerical framework and a baseline for assessing the influence of phase change. We then examine dewetting with phase change, with particular attention to the interfacial heat flux and interface-slope distributions near the contact line. Next, the DNS results are compared with the predictions of lubrication-based models commonly used to describe the phase-change-induced increase in interfacial slope. Finally, we investigate, in the presence of phase change, the effects of the equilibrium contact angle and slip length on the dewetting dynamics, and further examine the scale-separated interpretation of the interfacial-slope variation at a reduced slip length.

\subsection{Film dewetting without phase change}
\label{sec:3.1}

Isothermal dewetting at $\Delta T = 0$ is a classical problem \citep{gennes2004capillarity}, which typically exhibits two dynamical regimes, namely the viscous regime and the inertial regime. For the micrometre-scale liquid films considered here, the dewetting dynamics lies in the viscous regime, in which the film motion is governed primarily by the balance between capillary and viscous forces. Figure~\ref{fig:3-1-1}(a) shows the film profiles at $t = 0.4\,\mathrm{ms}$ for different equilibrium contact angles. As expected, the contact line recedes faster as $\theta_e$ increases. Under the action of capillary forces, the film front retracts and the displaced liquid accumulates near the receding front. Since the liquid in the unperturbed flat part of the film remains nearly motionless owing to the strong viscous resistance in such a thin layer, this accumulation leads to the formation of a rim, which is particularly pronounced for $\theta_e = 20^\circ$.

\begin{figure}[t]
	\centering
	\begin{subfigure}[t]{0.48\textwidth}
		\centering
		\includegraphics[width=\linewidth]{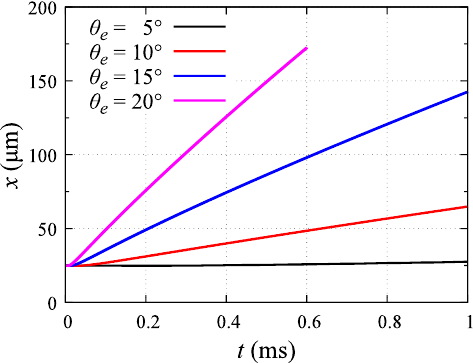}
		\subcaption{}
	\end{subfigure}
	\hfill
	\begin{subfigure}[t]{0.48\textwidth}
		\centering
		\includegraphics[width=\linewidth]{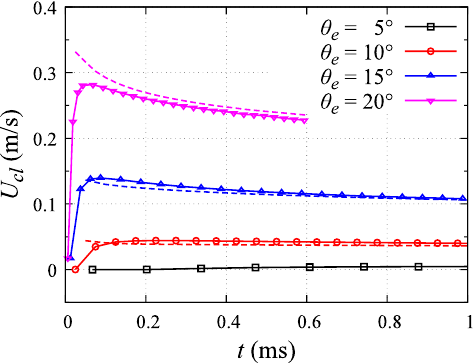}
		\subcaption{}
	\end{subfigure}
	\caption{Time evolution of (a) the contact-line position and (b) the contact-line speed. The dashed curves in (b) denote the contact-line speeds predicted by \eqref{eq:ca_theo}.}
	\label{fig:3-1-2}
\end{figure}

Figure~\ref{fig:3-1-2}(a) shows the time evolution of the contact-line position for different equilibrium contact angles. At $\theta_e = 5^\circ$, the contact line remains almost motionless for $t < 1\,\mathrm{ms}$ because of the very small contact angle. As $\theta_e$ increases, the contact line retracts more rapidly, consistent with the trend already observed in figure~\ref{fig:3-1-1}(a). Over most of the simulated time interval, the retraction is approximately linear with time, indicating that the contact line moves at an almost constant speed. To illustrate this more clearly, figure~\ref{fig:3-1-2}(b) shows the corresponding contact-line speed as a function of time. After a short initial acceleration stage, which is closely related to the initialisation of the simulation, the contact-line speed decreases slowly with time. This decrease is accompanied by an increase in the rim half-width $w$, defined in figure~\ref{fig:2-1-1}, and is consistent with the asymptotic analysis of isothermal dewetting at small contact angles by \citet{snoeijer2010asymptotic}, who derived the following expression for the contact-line speed:
\begin{equation}\label{eq:ca_theo}
	Ca = \frac{\mu_l U_{cl}}{\sigma} = \frac{\theta_{e}^{3}}{9}
	\left[
	\ln\!\left(
	\frac{4a}{3e}\,\theta_{e}\, Ca^{1/3}
	\frac{w^{2}}{\lambda_s h_{0}}
	\right)
	\right]^{-1},
\end{equation}
where $e$ is Euler's number and $a = 1.094$ is a constant. In figure~\ref{fig:3-1-2}(b), the numerical results are compared with \eqref{eq:ca_theo} for $\theta_e \geq 10^\circ$. It should be noted that the asymptotic analysis of \citet{snoeijer2010asymptotic} relies on the separation of three geometric scales, namely $\lambda_s \ll h_0 \ll h_r$, where $h_r$ is the rim height defined in figure~\ref{fig:2-1-1}. As can be seen from figure~\ref{fig:3-1-1}(a), the condition $h_0 \ll h_r$ is not  well satisfied in the present simulations. Nevertheless, except for the initial acceleration stage, the numerical results remain in good agreement with the theoretical prediction. For the case $\theta_e = 20^\circ$, the results for $t > 0.6\,\mathrm{ms}$ are not shown because the capillary wave generated by the receding contact line reaches the right boundary of the computational domain, causing oscillations in the contact-line speed due to the outflow boundary condition imposed there.

Next, we examine the interface-slope distribution in the near-contact-line region. Using asymptotic analysis, \citet{snoeijer2010asymptotic} derived a parametric solution for the rim profile of a dewetting film. Building on this solution, we derive an explicit expression for the interface slope $\theta$ in the near-contact-line region but outside the slip region:
\begin{equation}\label{solution0K-nonl}
	\theta^3 = \left( \theta_e^3 - 9\,Ca\ln\xi \right)
	\left(1 - \frac{3\,Ca}{\theta_e^3 - 9\,Ca\ln\xi}\right)^3,
\end{equation}
where $\xi=x\theta_e/(3\lambda_s)$ and $x$ is the horizontal distance from the contact line. The derivation of \eqref{solution0K-nonl} is given in Appendix~\ref{appB}. In the limit
\begin{equation}\label{eq:lin-cond}
	3\,Ca \ll \theta_e^3 - 9\,Ca\ln\xi,
\end{equation}
\eqref{solution0K-nonl} can be linearised as
\begin{equation}\label{solution0K-lin}
	\theta^3 = \theta_e^3 - 9\,Ca\ln\left(\frac{e\theta_e x}{3\lambda_s}\right),
\end{equation}
which has the same form as the classical Cox--Voinov law \citep{voinov1976hydrodynamics,Cox1986the}.

\begin{figure}[t]
	\centering
	\begin{subfigure}[t]{0.48\textwidth}
		\centering
		\includegraphics[width=\linewidth]{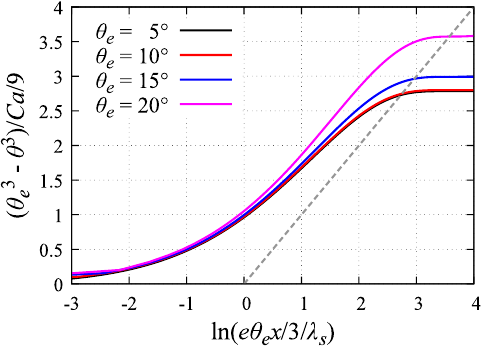}
		\subcaption{}
	\end{subfigure}
	\hfill
	\begin{subfigure}[t]{0.48\textwidth}
		\centering
		\includegraphics[width=\linewidth]{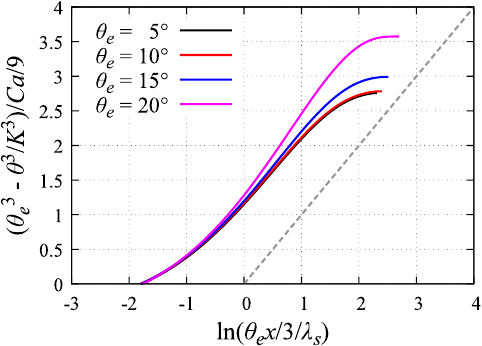}
		\subcaption{}
	\end{subfigure}
	\caption{Comparison of the numerical results for the interface-slope distribution with \eqref{solution0K-lin} in (a) and \eqref{solution0K-nonl} in (b). In (b), $K$ in the ordinate is defined by $K = 1 - 3\,Ca/(\theta_e^3 - 9\,Ca \ln \xi)$. The dashed line indicates equality between the abscissa and ordinate. The profiles are taken at $t=3.2$, $0.4$, $0.2$ and $0.2\,\mathrm{ms}$ for $\theta_e=5^\circ$, $10^\circ$, $15^\circ$ and $20^\circ$, respectively.}
	\label{fig:3-1-3}
\end{figure}

Figure~\ref{fig:3-1-3} compares the numerical results with the predictions of \eqref{solution0K-nonl} and \eqref{solution0K-lin}. The slope profiles are shown at representative times when the interface-slope distribution has reached a quasi-steady state, as illustrated in figure~\ref{fig:A-1}. In both panels, the data for different equilibrium contact angles collapse reasonably well, especially in the near-contact-line region, indicating a similar dependence of the interface slope on distance. However, when the numerical results are compared with the linearised expression \eqref{solution0K-lin} in figure~\ref{fig:3-1-3}(a), the curves deviate visibly from the dashed line of equality which corresponds to exact agreement with \eqref{solution0K-lin}, especially for smaller equilibrium contact angles. In figure~\ref{fig:3-1-3}(b), by contrast, comparison with the full expression \eqref{solution0K-nonl} shows a clear improvement in the slope: the curves become nearly parallel to the dashed line of equality, although a finite offset remains. In \eqref{solution0K-nonl}, the logarithmic argument can be written as $\xi=x/l_{\mathrm{m}}$, where $l_{\mathrm{m}}=3\lambda_s/\theta_e$ represents the  effective microscopic length scale related to the slip region. The remaining offset may thus arise from a shift in this microscopic length scale when compared with DNS, as well as higher-order corrections neglected in the derivation of \eqref{solution0K-nonl}. Nevertheless, \eqref{solution0K-nonl} provides a slightly better representation of the numerical results than the linearised counterpart, because the linearisation condition \eqref{eq:lin-cond} is not well satisfied when the equilibrium contact angle is very small.

Finally, we examine the apparent contact angle. Here, $\theta_a$ is defined by fitting a parabola to the rim profile using the rim height $h_r$ and its half-width $w$, as illustrated in figure~\ref{fig:2-1-1}(a). Based on the asymptotic analysis of \citet{snoeijer2010asymptotic}, two expressions for $\theta_a$ are obtained:
\begin{equation}\label{eq:app_inner}
	\theta_{a}^{3} = \theta_{e}^{3} - 9\,Ca \ln\!\left(\frac{2\,\theta_{e}w}{3\,\lambda_s}\right),
\end{equation}
\begin{equation}\label{eq:app_outer}
	\theta_{a}^{3} = 9\,Ca\ln\!\left(\frac{2a\,Ca^{1/3}w}{e\,h_0}\right).
\end{equation}
The first relation is obtained from the inner rim solution, where the relevant length scales are the macroscopic rim half-width $w$ and the microscopic slip length $\lambda_s$. In contrast, the second relation is obtained from the outer film solution, in which the relevant length scales are $w$ and the initial film thickness $h_0$.

\begin{figure}[t]
	\centering
	\begin{subfigure}[t]{0.495\textwidth}
		\centering
		\includegraphics[width=\linewidth]{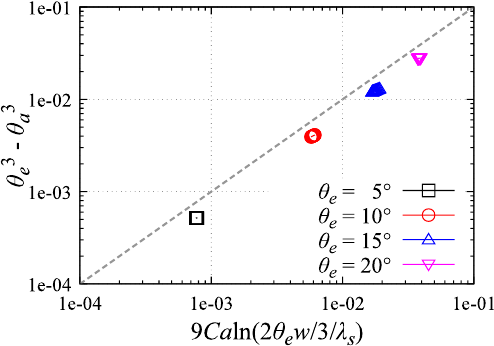}
		\subcaption{}
	\end{subfigure}
	\hfill
	\begin{subfigure}[t]{0.47\textwidth}
		\centering
		\includegraphics[width=\linewidth]{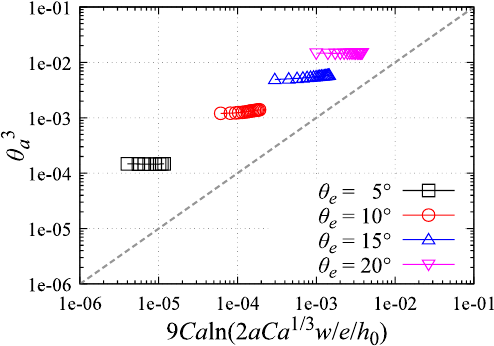}
		\subcaption{}
	\end{subfigure}
	\caption{Comparison of the numerical results for the apparent contact angle $\theta_a$ with the theoretical predictions at different times. Here, $\theta_a$ is determined from the parabola fitted using the rim height $h_r$ and half-width $w$, as illustrated in figure~\ref{fig:2-1-1}(a), namely, $\theta_a = 2h_r/w$. (a) Comparison with \eqref{eq:app_inner}. (b) Comparison with \eqref{eq:app_outer}.}	
	\label{fig:3-1-4}
\end{figure}

In figure~\ref{fig:3-1-4}, the simulated results for different equilibrium contact angles are compared with the predictions of \eqref{eq:app_inner} and \eqref{eq:app_outer}. During film dewetting, both the rim half-width $w$ and the contact-line speed, and hence $Ca$, vary with time. As a result, for each equilibrium contact angle considered here, the data form a series of points corresponding to different time instants. As shown in figure~\ref{fig:3-1-4}(a), the numerical results agree reasonably well with \eqref{eq:app_inner}, whereas the agreement with \eqref{eq:app_outer} in figure~\ref{fig:3-1-4}(b) is less satisfactory. This difference can be attributed to the degree of scale separation achieved in the present simulations. The microscopic slip length $\lambda_s$ is well separated from the macroscopic rim half-width $w$, namely $\lambda_s\ll w$, while the separation between the two macroscopic length scales $w$ and $h_0$ remains limited as shown in figure~\ref{fig:3-1-1}(a). Consequently, \eqref{eq:app_outer}, which is an asymptotic result requiring a clear separation between $w$ and $h_0$, is not reproduced as well as \eqref{eq:app_inner} in the present simulations.

\subsection{Film dewetting with phase change}
\label{sec:3.2}

We next examine how phase change influences the dewetting dynamics. The isothermal results in \S\ref{sec:3.1} show that, among the equilibrium contact angles considered here, the case $\theta_e=5^{\circ}$ exhibits the slowest capillary-driven dewetting. It therefore provides a sensitive configuration for isolating the effect of evaporation. Unless otherwise stated, we focus in this section on $\theta_e=5^{\circ}$, while varying the wall superheat over the range $\Delta T=0$--$16\,\mathrm{K}$ and the accommodation coefficient over the range $f_{ac}=0.05$--$0.4$. These two parameters control the intensity of phase change: increasing $\Delta T$ increases the temperature difference between the wall and liquid--vapour interface, whereas increasing $f_{ac}$ reduces the interfacial thermal resistance. We first describe the resulting changes in film morphology and retraction speed, and then analyse the associated interfacial heat flux, interface-slope distribution and apparent contact angle.

\begin{figure}[t]
	\centering
	
	\begin{subfigure}{\linewidth}
		\centering
		\includegraphics[width=\linewidth]{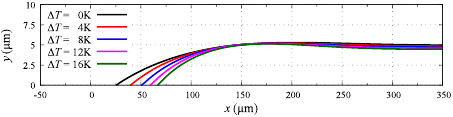}
		\caption{}\label{fig:3-2-1a}
	\end{subfigure}
	\vspace{0.4em}
	\begin{subfigure}{\linewidth}
		\centering
		\includegraphics[width=\linewidth]{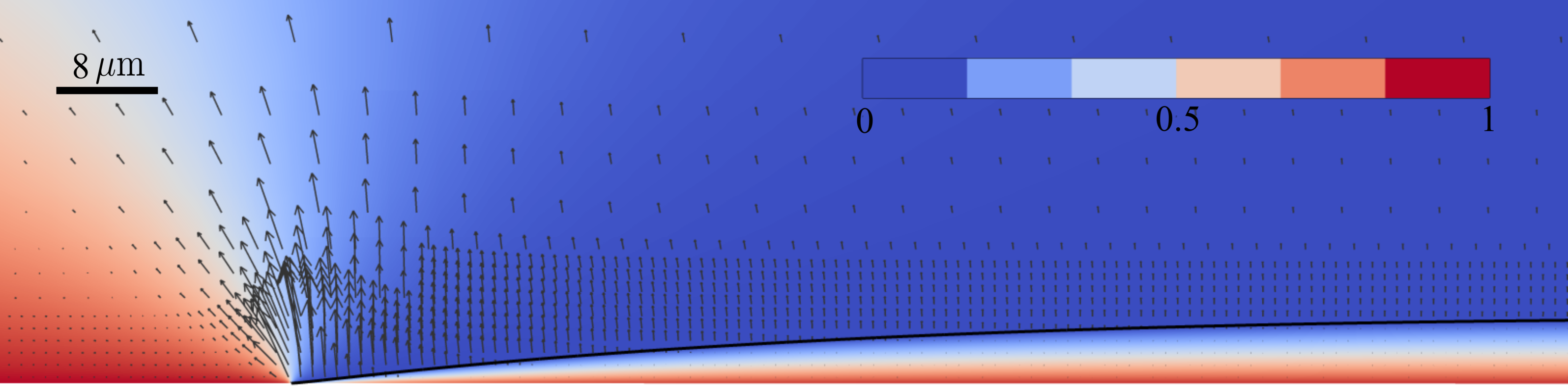}
		\caption{}\label{fig:3-2-1b}
	\end{subfigure}
	\caption{Film profiles, temperature contours and velocity vectors at $\theta_e=5^{\circ}$ and $f_{ac}=0.4$. (a) Film profiles at $t=0.5\,\mathrm{ms}$ for different wall superheats $\Delta T$; the vertical scale is enlarged for clarity. (b) Contours of the dimensionless temperature $(T-T_s)/\Delta T$ and velocity vectors near the contact line for $\Delta T=8\,\mathrm{K}$ at the same time instant.}
	
	\label{fig:3-2-1}
\end{figure}

Figure~\ref{fig:3-2-1}(a) compares the film profiles at $t=0.5\,\mathrm{ms}$ for different wall superheats at $f_{ac}=0.4$. Two consequences of phase change are apparent. First, evaporation accelerates the contact-line recession, as shown by the larger displacement of the contact line from its initial position at $x = 25\,\mathrm{\mu m}$. Secondly, the film profile becomes steeper near the receding front, corresponding to an increase in the apparent contact angle at the macroscopic scale, even though the same equilibrium contact angle, $\theta_e=5^{\circ}$, is imposed in all cases. Both the contact-line displacement and the apparent contact angle increase with wall superheat, indicating that evaporation progressively enhances the dewetting dynamics.

Figure~\ref{fig:3-2-1}(b) shows contours of the dimensionless temperature, $(T-T_s)/\Delta T$, together with velocity vectors in the near-contact-line region for the representative case with $f_{ac}=0.4$ and $\Delta T=8\,\mathrm{K}$. In contrast to panel (a), the vertical scale is not enlarged. The temperature field in the thin liquid film varies mainly in the wall-normal direction, indicating that heat transfer through the film is governed primarily by one-dimensional conduction. The velocity field also reflects the effect of phase change. As indicated by the velocity jump relation \eqref{eq:jump-u}, for a given interfacial mass flux, the normal velocity of the vapour relative to the interface is much larger than that of the liquid because of $\rho_v \ll \rho_l$. Consequently, the velocity magnitude in the liquid film is much smaller than that in the vapour. Moreover, the vapour velocity increases markedly towards the contact line, where the liquid film becomes thinner and the local heat flux, and hence the evaporation rate, is enhanced. This localized increase in evaporation rate near the contact line is consistent with the heat-flux distributions analysed below.

\begin{figure}[t]
	\centering
	\begin{subfigure}[t]{0.48\textwidth}
		\centering
		\includegraphics[width=\linewidth]{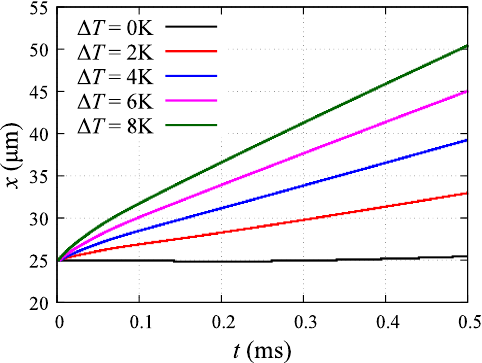}
		\subcaption{}
	\end{subfigure}
	\hfill
	\begin{subfigure}[t]{0.48\textwidth}
		\centering
		\includegraphics[width=\linewidth]{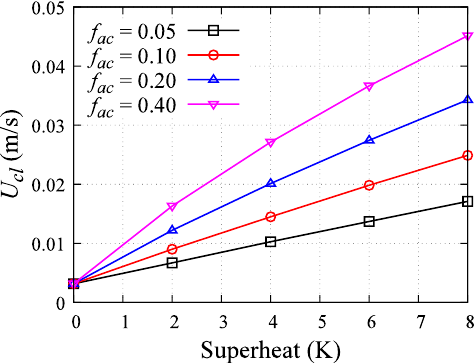}
		\subcaption{}
	\end{subfigure}
	\caption{Contact-line motion at $\theta_e=5^{\circ}$. (a) Time evolution of the contact-line position for $f_{ac}=0.4$ at different wall superheats $\Delta T$. (b) Contact-line speed at $t=0.5\,\mathrm{ms}$ as a function of wall superheat for different accommodation coefficients $f_{ac}$.}
	\label{fig:3-2-2}
\end{figure}

Figure~\ref{fig:3-2-2}(a) shows the time evolution of the contact-line position for different wall superheats at $f_{ac}=0.4$. After a short initial transient, the contact line recedes at an almost constant speed. This speed increases with wall superheat, consistent with the larger contact-line displacement observed in figure~\ref{fig:3-2-1}(a). To quantify this trend, figure~\ref{fig:3-2-2}(b) compares the contact-line speed at $t=0.5\,\mathrm{ms}$ for different wall superheats and accommodation coefficients. For each value of $f_{ac}$, the receding speed increases with $\Delta T$, while at any fixed $\Delta T$, the dewetting is faster for larger $f_{ac}$. This behaviour is consistent with the enhanced evaporation rate resulting from either increasing the wall superheat or reducing the interfacial thermal resistance.

\subsubsection{Interfacial heat flux}

\begin{figure}[t]
	\centering
	\begin{subfigure}[t]{0.48\textwidth}
		\centering
		\includegraphics[width=\linewidth]{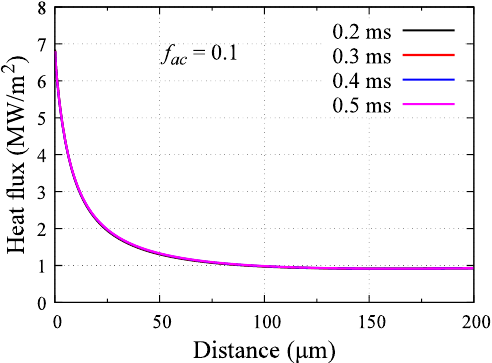}
		\subcaption{}
	\end{subfigure}
	\hfill
	\begin{subfigure}[t]{0.48\textwidth}
		\centering
		\includegraphics[width=\linewidth]{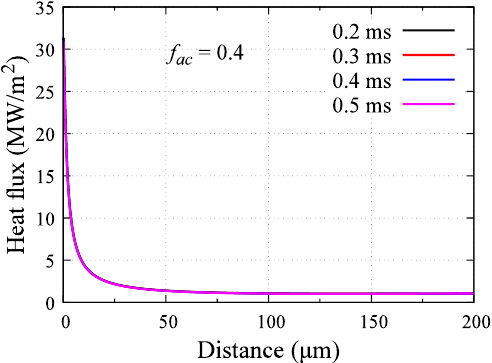}
		\subcaption{}
	\end{subfigure}
	\begin{subfigure}[t]{0.48\textwidth}
		\centering
		\includegraphics[width=\linewidth]{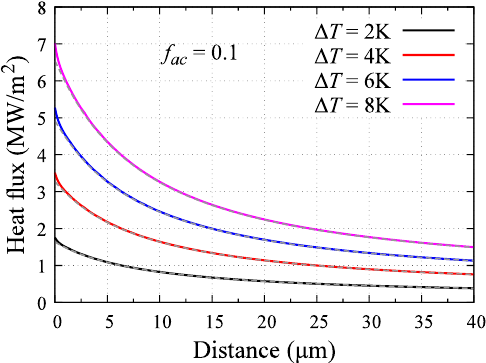}
		\subcaption{}
	\end{subfigure}
	\hfill
	\begin{subfigure}[t]{0.48\textwidth}
		\centering
		\includegraphics[width=\linewidth]{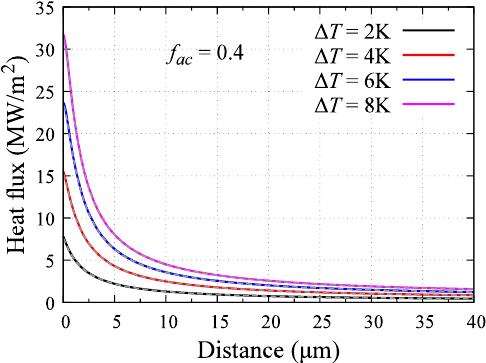}
		\subcaption{}
	\end{subfigure}
	\caption{Interfacial heat flux as a function of the distance from the instantaneous contact line for $\theta_e=5^{\circ}$. (a,b) Temporal evolution at $\Delta T=8\,\mathrm{K}$ for (a) $f_{ac}=0.1$ and (b) $f_{ac}=0.4$. (c,d) Dependence on wall superheat at $t=0.4\,\mathrm{ms}$ for (c) $f_{ac}=0.1$ and (d) $f_{ac}=0.4$. The grey dashed curves in (c,d) denote the prediction of \eqref{eq:heat-flux-analy}.}
	\label{fig:3-2-3}
\end{figure}

The preceding results show that evaporation accelerates the contact-line recession and modifies the film shape. Since the phase-change mass flux at the liquid--vapour interface is determined by the interfacial energy balance \eqref{eq:jump-q}, we first examine the corresponding interfacial heat-flux distribution. Figure~\ref{fig:3-2-3} shows the heat flux as a function of the distance from the instantaneous contact line for two representative accommodation coefficients, $f_{ac}=0.1$ and $f_{ac}=0.4$. Figures~\ref{fig:3-2-3}(a,b) show the temporal evolution of the heat-flux distribution at $\Delta T=8\,\mathrm{K}$. Over the time interval considered, the profiles nearly collapse when plotted in the contact-line-fixed frame, indicating that the heat-flux distribution has reached a quasi-steady state. This behaviour is consistent with the temperature field shown in figure~\ref{fig:3-2-1}(b), where heat transfer through the thin liquid film is mainly wall-normal. Figures~\ref{fig:3-2-3}(c,d) compare the interfacial heat flux for different wall superheats. As expected, increasing $\Delta T$ increases the heat-flux magnitude and hence strengthens the local evaporation rate. At a fixed wall superheat, the case $f_{ac}=0.4$ exhibits both a larger heat flux magnitude and a more rapid decay away from the contact line than the case $f_{ac}=0.1$. This behaviour results from the smaller interfacial thermal resistance associated with the larger accommodation coefficient, which not only increases the maximum heat flux but also makes the heat-flux distribution more concentrated near the contact line.

To further test the wall-normal conduction picture suggested by figure~\ref{fig:3-2-1}(b), we compare the DNS results with the heat flux predicted by a local one-dimensional conduction model. The corresponding prediction, shown by the dashed curves in figures~\ref{fig:3-2-3}(c,d), is
\begin{equation}
	\label{eq:heat-flux-analy}
	q_l = \frac{T_w - T_s}{h/k_l + R_{\Gamma}},
\end{equation}
where $h$ is the local film thickness. This expression follows directly from the heat-flux balances in \eqref{eq:flux1} and \eqref{eq:flux2}. Overall, \eqref{eq:heat-flux-analy} agrees well with the numerical results, except for a minor deviation observed only very close to the contact line, where the simulated heat flux is slightly larger than the predicted value. This discrepancy may be associated with the first-order heat-flux evaluation used in the first few cells adjacent to the contact line, as indicated by \eqref{eq:one-side}. Nevertheless, the agreement between the DNS results and \eqref{eq:heat-flux-analy} supports the use of a one-dimensional heat conduction approximation within the liquid film, as well as the numerical treatment in which the interfacial thermal resistance is imposed through an equivalent boundary condition at the liquid--solid interface.

\begin{figure}[t]
	\centering
	\begin{subfigure}[t]{0.48\textwidth}
		\centering
		\includegraphics[width=\linewidth]{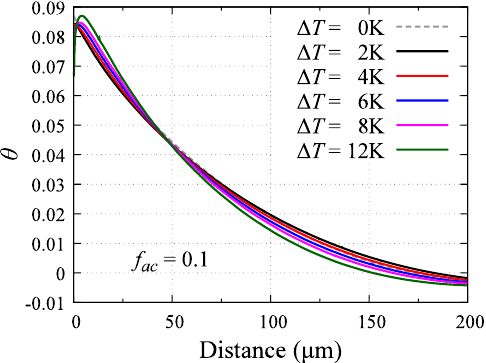}
		\subcaption{}
	\end{subfigure}
	\hfill
	\begin{subfigure}[t]{0.48\textwidth}
		\centering
		\includegraphics[width=\linewidth]{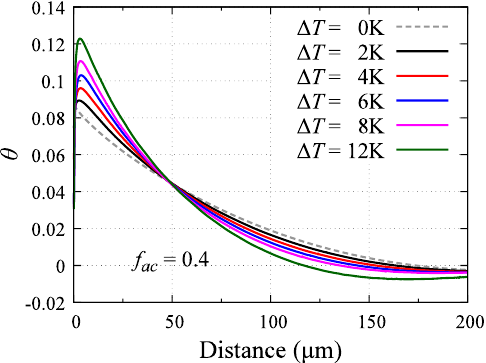}
		\subcaption{}
	\end{subfigure}
	\begin{subfigure}[t]{0.48\textwidth}
		\centering
		\includegraphics[width=\linewidth]{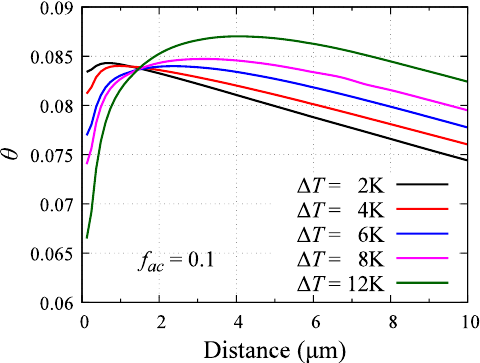}
		\subcaption{}
	\end{subfigure}
	\hfill
	\begin{subfigure}[t]{0.48\textwidth}
		\centering
		\includegraphics[width=\linewidth]{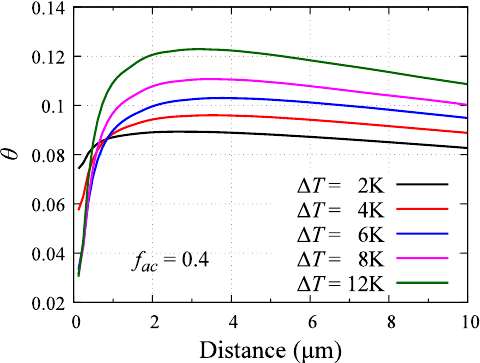}
		\subcaption{}
	\end{subfigure}
	\caption{Interfacial slope versus the distance from the instantaneous contact line for $\theta_e=5^{\circ}$, taken after the slope distribution has reached a quasi-steady state. (a,b) Dependence on wall superheat for (a) $f_{ac}=0.1$ and (b) $f_{ac}=0.4$. Panels (c,d) show enlarged views of the near-contact-line regions in (a,b), respectively.}
	\label{fig:3-2-4}
\end{figure}

\subsubsection{Interface slope distribution}
\label{sec:slope}

Next, we examine the interfacial-slope distribution as a function of the distance from the contact line. Figure~\ref{fig:3-2-4} compares the numerical results obtained at different wall superheats for two representative accommodation coefficients, $f_{ac}=0.1$ and $f_{ac}=0.4$. In each case, the profile is taken after the slope distribution has reached a quasi-steady state, as demonstrated in figure~\ref{fig:A-4}. For reference, the isothermal result, corresponding to $\Delta T=0$, is also included. Figures~\ref{fig:3-2-4}(c,d) show enlarged views of the near-contact-line regions in figures~\ref{fig:3-2-4}(a,b), respectively. Note that the slope in the contact-line cell is excluded from the plotted distributions because the interface reconstruction in this cell is directly constrained by the imposed equilibrium contact angle $\theta_e$ \citep{han2026applying}. Compared with the isothermal case, phase change modifies the interfacial-slope distribution in two ways. First, the interfacial slope resolved by the DNS immediately adjacent to the contact line decreases as the wall superheat increases, although the prescribed equilibrium contact angle $\theta_e$ remains unchanged. This decrease is particularly visible in the enlarged view for $f_{ac}=0.1$, shown in figure~\ref{fig:3-2-4}(c). With sufficiently strong phase change, for example in the cases of $\Delta T>6\,\mathrm{K}$ and $f_{ac}=0.4$, the slope in the first few cells becomes very small, indicating that the interface there is nearly parallel to the solid wall. Secondly, a local maximum of the interfacial slope develops near the contact line, and its magnitude increases with wall superheat. This effect is more pronounced for $f_{ac}=0.4$, as shown in figure~\ref{fig:3-2-4}(d), consistent with the stronger and more localized heat flux in figures~\ref{fig:3-2-3}(c,d). For $f_{ac}=0.1$, the same effect becomes evident only when the wall superheat is sufficiently large. These two changes together give rise to a near-contact-line region in which the interfacial slope increases with distance from the contact line. Such a phase-change-induced increase in slope is consistent with the prediction of micro-region models for evaporating contact lines \citep{mathieu2003}.

\begin{figure}[t]
	\centering
	\begin{subfigure}[t]{\textwidth}
		\centering
		\includegraphics[width=\linewidth]{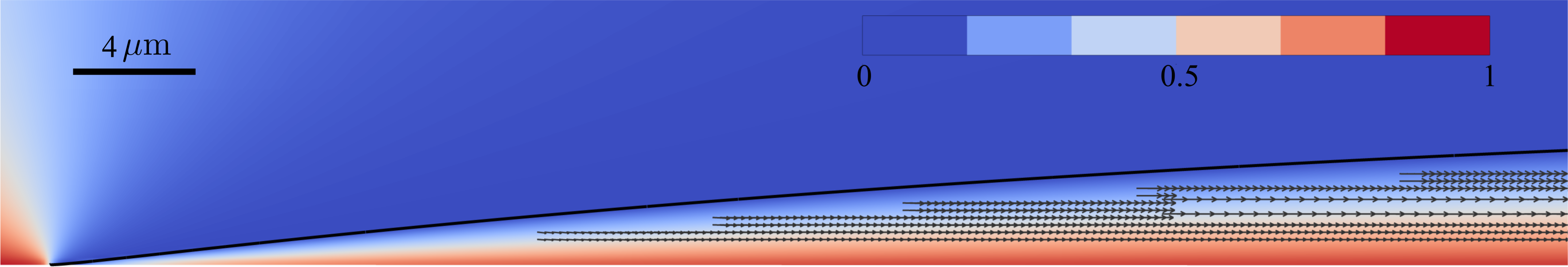}
		\subcaption{}
	\end{subfigure}
	\begin{subfigure}[t]{\textwidth}
		\centering
		\includegraphics[width=\linewidth]{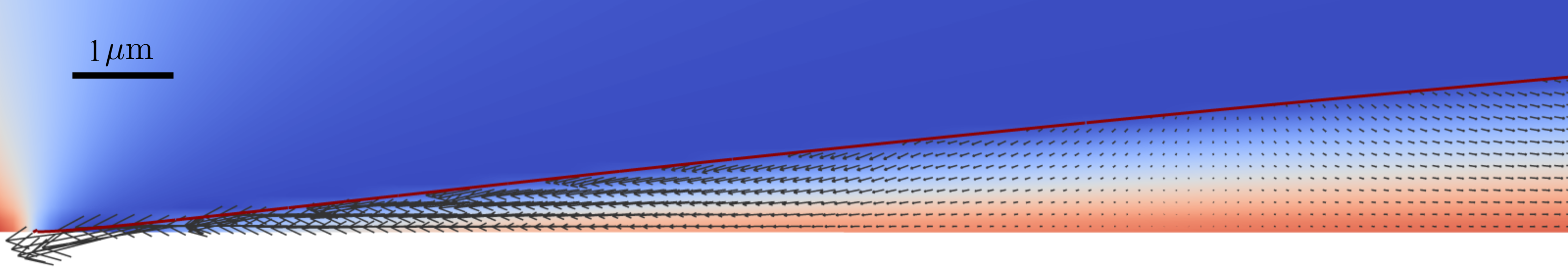}
		\subcaption{}
	\end{subfigure}
	\caption{Liquid velocity field near the contact line for $\theta_e=5^{\circ}$, $f_{ac}=0.4$ and $\Delta T=8\,\mathrm{K}$ at $t=0.4\,\mathrm{ms}$. The vectors show the liquid velocity $\boldsymbol{u}_l$, superimposed on contours of the dimensionless temperature $(T-T_s)/\Delta T$. Panel (b) shows an enlarged view of the near-contact-line region in panel (a).}
	\label{fig:3-2-9}
\end{figure}

The near-contact-line increase in the interfacial slope helps explain the increase in the apparent contact angle observed in figure~\ref{fig:3-2-1}(a). This local slope increase is associated with intense evaporation in the vicinity of the contact line. Although this effect is already suggested by the temperature and velocity fields shown in figure~\ref{fig:3-2-1}(b), the liquid motion is difficult to identify there because the phase-change-induced velocity jump, together with the large density ratio, makes the vapour velocity much larger than the liquid velocity. To isolate the flow inside the liquid film, figure~\ref{fig:3-2-9} shows the liquid velocity field for the representative case with $f_{ac}=0.4$ and $\Delta T=8\,\mathrm{K}$. Away from the contact line, the liquid velocity is predominantly parallel to the solid wall, as shown in figure~\ref{fig:3-2-9}(a), consistent with the lubrication approximation usually employed to describe dewetting at small equilibrium contact angles. In contrast, in the immediate vicinity of the contact line, the velocity vectors turn towards the contact line, as can be seen in figure~\ref{fig:3-2-9}(b). This reversal reflects the evaporation-induced liquid flow that supplies mass to the region of large evaporative flux. Thus, the liquid flow exhibits a transition from evaporation-driven motion near the contact line to capillary-driven film retraction farther away from it. A similar competition between evaporation-induced flow and contact-line recession was discussed by \citet{berteloot2008evaporation} using a lubrication model for evaporative coating. Although their evaporation law is diffusion-limited, unlike the heat-conduction-dominated phase change considered here, their model also predicts a stagnation point where the evaporation-induced flow balances the flow associated with contact-line recession. The velocity reversal observed in figure~\ref{fig:3-2-9}(b) can be interpreted as the DNS counterpart of this transition. Very close to the contact line, figure~\ref{fig:3-2-9}(b) also shows a downward wall-normal component in the interfacial cells adjacent to the solid wall. This component may be related to the local normal motion of the liquid--vapour interface induced by phase change.

The preceding results suggest a transition from an evaporation-dominated micro-region very close to the contact line to a capillary-driven region farther away. This observation naturally raises the question of whether the effects of phase change and contact-line motion on the interfacial slope can be treated separately. Following this idea, \citet{zhang2022dewetting} proposed that the interface-slope distribution outside the micro-region can be described by a Cox--Voinov-type relation,
\begin{equation}
	\label{eq:zhang}
	\theta^3 = \theta_V^3 - 9\,Ca\ln\frac{x}{l_V},
\end{equation}
where $\theta_V$ is the Voinov angle and $l_V$ is the corresponding Voinov length. In this framework, $\theta_V$ represents the interfacial slope at the outer end of the micro-region, which differs from the equilibrium contact angle $\theta_e$ due to phase change. The length $l_V$ may be regarded as an effective length scale associated with the micro-region, at which the Cox--Voinov relation \eqref{eq:zhang} gives $\theta=\theta_V$. The key assumption underlying \eqref{eq:zhang} is that phase change and contact-line motion affect the slope distribution on separated length scales: the effect of phase change takes place inside the micro-region and is represented by the Voinov angle $\theta_V$, whereas contact-line motion produces the logarithmic bending outside the micro-region. This scale-separated interpretation is assessed below using the present DNS results.

\begin{figure}[t]
	\centering
	\begin{subfigure}[t]{0.48\textwidth}
		\centering
		\includegraphics[width=\linewidth]{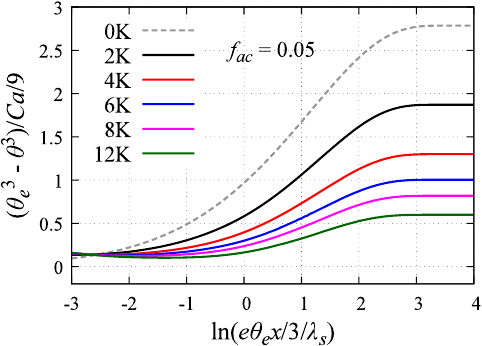}
		\subcaption{}
	\end{subfigure}
	\hfill
	\begin{subfigure}[t]{0.48\textwidth}
		\centering
		\includegraphics[width=\linewidth]{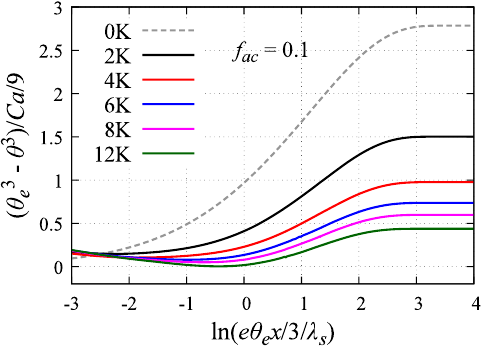}
		\subcaption{}
	\end{subfigure}
	\caption{Interfacial-slope distributions replotted using the Cox--Voinov scaling implied by \eqref{eq:zhang} for $\theta_e=5^{\circ}$. Results are shown for (a) $f_{ac}=0.05$ and (b) $f_{ac}=0.1$. The profiles are taken after the slope distribution has reached a quasi-steady state. The dashed curve denotes the isothermal-dewetting result at $\theta_e=5^{\circ}$.}
	\label{fig:3-2-5}
\end{figure}

In figure~\ref{fig:3-2-5}, the slope profiles for the two smallest accommodation coefficients, $f_{ac}=0.05$ and 0.1, are replotted using the scaling implied by \eqref{eq:zhang}, where we assume $\theta_V=\theta_e$ and $l_V=3\lambda_s/(e\theta_e)$. If the Cox--Voinov-type description were sufficient in the intermediate region, the transformed data would remain approximately parallel to the isothermal curve (figure~\ref{fig:3-1-3}(a)). However, this is not observed. Instead, the curves corresponding to different wall superheats exhibit slopes that are smaller than that of the isothermal curve and decrease with increasing $\Delta T$. Over the range $0\le\ln(e\theta_e x/3\lambda_s)\le2$, linear fits give slopes decreasing from about 0.55 at $\Delta T=2$\,K to 0.20 at $\Delta T=12$\,K for $f_{ac}=0.05$, and from about 0.47 to 0.20 for $f_{ac}=0.1$, compared with about 0.75 for the isothermal curve. It should be emphasized that whatever values are assigned to $\theta_V$ and $l_V$, they can only shift the curves in this representation, but cannot alter their slopes. The same conclusion holds for $f_{ac}=0.4$ (see figure~\ref{fig:A-5}). We therefore conclude that, over the range of length scales and parameters examined in the present DNS, the evaporation-induced changes in the interfacial slope outside the micro-region cannot be captured solely by adjusting the Voinov angle and length in \eqref{eq:zhang}.

To describe the interface-slope distributions observed in the present simulations, we return to the steady lubrication equation. This description is appropriate here because the interface slopes remain small, as shown in figure~\ref{fig:3-2-4}, and the slope profiles have reached a quasi-steady state in a frame moving with the contact line. In this moving frame, the lubrication equation can be written as \citep{hocking1995contact}
\begin{equation}\label{lubrication}
	\frac{\partial}{\partial \hat{x}}\left[\left(\frac{1}{3}\hat{h}+\hat{\lambda}_s\right)\hat{h}^2\,\frac{\partial^3 \hat{h}}{\partial \hat{x}^3} - \frac{Ca}{\theta_e^3}\hat{h}\right]
	+\frac{Ca_e}{\theta_e^4}\frac{1}{1+\hat{h}} =0,
\end{equation}
where $\hat{h}$ and $\hat{x}$ denote the dimensionless film thickness and distance from the contact line, obtained by rescaling the corresponding dimensional quantities by $l_R$ and $l_R/\theta_e$, respectively, and $\hat{\lambda}_s=\lambda_s/l_R$ is the dimensionless slip length. The corresponding dimensional form is given by \eqref{eq:app_lubrication}. The term proportional to $Ca_e$ represents the local evaporation rate resulting from the one-dimensional heat-conduction model. Here, $Ca_e$ is the evaporative capillary number \citep{morris2001contact}, defined as
\begin{equation}
	Ca_e = \frac{\mu_l}{\sigma}\frac{k_l \Delta T}{\rho_l L_e l_R},
\end{equation}
which is a capillary number based on the characteristic velocity induced by phase change near the contact line. Equation~\eqref{lubrication} shows that, for the receding configuration considered here, contact-line motion, represented by the hydrodynamic term proportional to $Ca$, and phase change, represented by the evaporation term proportional to $Ca_e$, act in opposition in shaping the interface-slope distribution. The former tends to reduce the interface slope with increasing distance from the contact line, as already verified for isothermal dewetting in figure~\ref{fig:3-1-3}, whereas the latter tends to increase the slope near the contact line, consistent with the behaviour observed in figure~\ref{fig:3-2-4}.

\begin{figure}[t]
	\centering
	\begin{subfigure}[t]{0.48\textwidth}
		\centering
		\includegraphics[width=\linewidth]{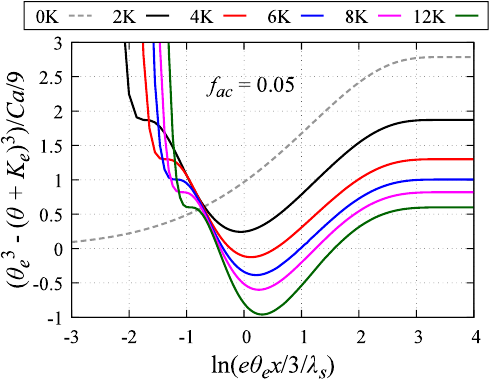}
		\subcaption{}
	\end{subfigure}
	\hfill
	\begin{subfigure}[t]{0.48\textwidth}
		\centering
		\includegraphics[width=\linewidth]{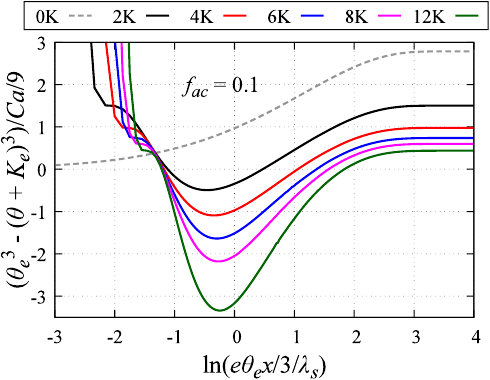}
		\subcaption{}
	\end{subfigure}
	\caption{Interface-slope distributions compared with \eqref{solution-large-x} for $\theta_e=5^{\circ}$. Results are shown for (a) $f_{ac}=0.05$ and (b) $f_{ac}=0.1$, with $K_e = Ca_e l_R \left[1.5\ln\left(\theta_e x/l_R\right)+2.43\right]/\left(\theta_e^4 x\right)$ used in the ordinate.  Each profile is taken after the slope distribution has reached a quasi-steady state. The dashed curves denote the isothermal-dewetting result at $\theta_e=5^{\circ}$.}
	\label{fig:3-2-6}
\end{figure}

The relative importance of the two contributions can be estimated by integrating \eqref{lubrication} once from the contact line, where the liquid flux vanishes. Approximating the interface by a wedge, $\hat{h}\simeq\hat{x}$, the resulting hydrodynamic and evaporative flux terms are $(Ca/\theta_e^3)\hat{h}$ and $(Ca_e/\theta_e^4)\ln(1+\hat{h})$, respectively, and their ratio is
\begin{equation}
	R(\hat{h})\simeq\varepsilon\,\frac{\ln(1+\hat{h})}{\hat{h}},
	\qquad
	\varepsilon=\frac{Ca_e}{\theta_e\,Ca}.
	\label{eq:ratio}
\end{equation}
Close to the contact line ($\hat{h}\ll1$), the heat flux saturates at $\Delta T/R_\Gamma$ and $R\to\varepsilon$; farther away ($\hat{h}\gg1$), $R$ decays only as $\ln\hat{h}/\hat{h}$. 
The scale-separated description underlying \eqref{eq:zhang} therefore rests on two conditions. First, neglecting contact-line motion within the micro-region, as is usually assumed when $\theta_V$ is predicted from a micro-region model, requires evaporation to dominate there, i.e.\ $\varepsilon\gg1$.  Second, a Cox--Voinov region can develop outside the micro-region only if $R\ll1$ over a range of $\hat{h}$ well below the macroscopic scale $h_0/l_R$. Since $R$ decreases monotonically with $\hat{h}$, this requires at least $R(h_0/l_R)\ll1$, which is possible only in films much thicker than $l_R$. When evaluated with the contact-line speeds measured in the DNS, $\varepsilon$ ranges from about 0.5 to 1.5 for the cases shown in figure~\ref{fig:3-2-5} and reaches about 4 for $f_{ac}=0.4$ (figure~\ref{fig:A-5}), so that $\varepsilon = O(1)$, while $h_0/l_R \approx 3$--$30$ (figure~\ref{fig:2-1-2}) gives $R(h_0/l_R) \approx 0.2$--$0.5$ in all cases. The violation of the second condition explains why the slope distributions shown in figures~\ref{fig:3-2-5} and~\ref{fig:A-5} cannot be collapsed by adjusting $\theta_V$ and $l_V$, whereas the consequence of $\varepsilon = O(1)$ within the micro-region is examined in \S\ref{sec:3.5}. By contrast, for the $50\,\mu$m-thick water films considered by \citet{zhang2022dewetting}, $h_0/l_R\approx10^3$, and $R(h_0/l_R)$ is more than an order of magnitude smaller for the same $\varepsilon$.

%so that the two contributions have the same dependence on $\hat{h}$

Although the two contributions are of comparable magnitude ($R=O(1)$), each only weakly bends the interface when $Ca/\theta_e^3\ll1$ and $Ca_e/\theta_e^4\ll1$, so that the interface deviates only slowly from a wedge. In the following, we consider this weak-phase-change regime and treat the evaporation term as a first-order perturbation of the Cox--Voinov solution, expanding $\hat{h}$ as
\begin{equation}
	\hat{h}= \hat{h}_0 + \frac{Ca_e}{\theta_e^4}\hat{h}_1 + \cdots
\end{equation}
In the region where the film thickness is larger than both the thermal-resistance length and the slip length, namely $\hat{h}\gg 1$ and $\hat{h}\gg \hat{\lambda}_s$, \eqref{lubrication} reduces to
\begin{equation}
	\frac{\partial}{\partial \hat{x}}\left[\frac{1}{3}\hat{h}^3\,\frac{\partial^3 \hat{h}}{\partial \hat{x}^3} - \frac{Ca}{\theta_e^3}\hat{h}\right]
	+\frac{Ca_e}{\theta_e^4}\frac{1}{\hat{h}} =0.
\end{equation}
At leading order, $\hat{h}_0$ satisfies
\begin{equation}
	\frac{\partial^3 \hat{h}_0}{\partial \hat{x}^3} = \frac{3\,Ca/\theta_e^3}{\hat{h}_0^2},
\end{equation}
whose solution has the Cox--Voinov form corresponding to \eqref{solution0K-lin}
\begin{equation}\label{eq:h0}
	\hat{\theta}_0^3 = 1 - \frac{9\,Ca}{\theta_e^3}\ln\left(\frac{e l_R\hat{x}}{3\lambda_s}\right),
\end{equation}
where $\hat{\theta}_0 = \partial \hat{h}_0/\partial \hat{x}$. The first-order correction $\hat{h}_1$ then satisfies,  to leading order in $Ca/\theta_e^3$,
%\begin{equation}
%	\frac{\partial}{\partial \hat{x}}\left[\frac{1}{3}\hat{h}_0^3\frac{\partial^3 \hat{h}_1}{\partial \hat{x}^3} + \frac{2\,Ca}{\theta_e^3}\hat{h}_1\right] + \frac{1}{\hat{h}_0} = 0
%\end{equation}
\begin{equation}
	\frac{\partial}{\partial \hat{x}}\left(\frac{1}{3}\hat{h}_0^3\frac{\partial^3 \hat{h}_1}{\partial \hat{x}^3}\right) + \frac{1}{\hat{h}_0} = 0.
\end{equation}
%\begin{equation}
%	\frac{\partial}{\partial \hat{x}}\left[\frac{1}{3}\hat{x}^3\frac{\partial^3  \hat{h}_1}{\partial \hat{x}^3}\right] + \frac{1}{\hat{x}} = 0
%\end{equation}
With the approximation $\hat{h}_0\simeq \hat{x}$, the corresponding slope correction is obtained as
\begin{equation}\label{eq:h1}
	\hat{\theta}_1 = \frac{\partial\hat{h}_1}{\partial\hat{x}}= \frac{C_1}{\hat{x}} -\frac{3}{2}\frac{\ln \hat{x}}{\hat{x}},
\end{equation}
where $C_1$ is an integration constant. This constant cannot be determined within the present large-$\hat{h}$ expansion, because the solution becomes singular as $\hat{x}\to 0$ and therefore depends on matching to the inner region. Combining \eqref{eq:h0} and \eqref{eq:h1}, and returning to dimensional variables, we obtain
\begin{equation} \label{solution-large-x}
	\theta = \left[\theta_e^3 - 9\,Ca\ln\left(\frac{e\theta_e x}{3\lambda_s}\right)\right]^{1/3} 
	- \frac{Ca_e l_R}{\theta_e^4 x}\left[\frac{3}{2}\ln\left(\frac{\theta_e x}{l_R}\right) - C_1\right].
\end{equation}

The solution \eqref{solution-large-x} can be interpreted as the Cox--Voinov solution \eqref{solution0K-lin} supplemented by a first-order correction due to evaporation. We compare this expression with the numerical results for the cases with small accommodation coefficients, namely $f_{ac}=0.05$ and $f_{ac}=0.1$, for which $Ca_e/\theta_e^4$ remains relatively small, ranging from about 0.03 to 0.2 for $f_{ac}=0.05$ and from about 0.06 to 0.4 for $f_{ac}=0.1$, in the range of $\Delta T=2$--$12$\,K. The results are shown in figure~\ref{fig:3-2-6}. The integration constant in \eqref{solution-large-x} is fixed at $C_1=-2.43$, obtained by fitting \eqref{solution-large-x} to the DNS result, and the same value is used for all cases considered. The isothermal-dewetting result at $\theta_e=5^{\circ}$ is also plotted for comparison. For $f_{ac}=0.05$, including the first-order evaporation correction makes the curves for different wall superheats nearly parallel to the isothermal result in the intermediate region, in contrast to the decreasing logarithmic slopes obtained without this correction (see figure~\ref{fig:3-2-5}(a)). A similar behaviour is observed for $f_{ac}=0.1$ when $\Delta T \leq 4\,\mathrm{K}$. At larger wall superheats, however, visible deviations remain, indicating that the weak-phase-change assumption underlying \eqref{solution-large-x} is no longer fully satisfied.

\begin{figure}[t]
	\centering
	\begin{subfigure}[t]{0.46\textwidth}
		\centering
		\includegraphics[width=\linewidth]{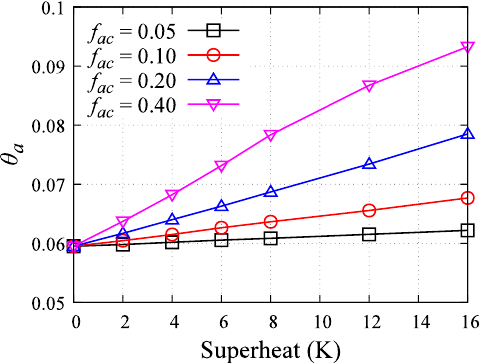}
		\subcaption{}
	\end{subfigure}
	\hfill
	\begin{subfigure}[t]{0.51\textwidth}
		\centering
		\includegraphics[width=\linewidth]{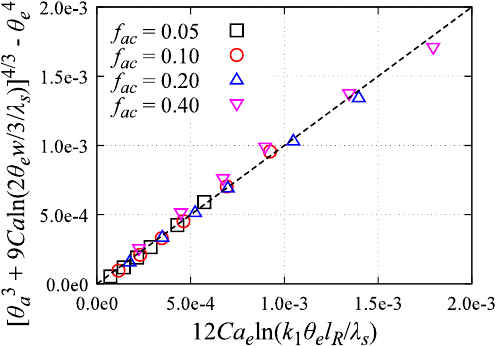}
		\subcaption{}
	\end{subfigure}
	\caption{Apparent contact angle $\theta_a$ for $\theta_e=5^{\circ}$, defined as $2h_r/w$. (a) Variation of $\theta_a$ with wall superheat $\Delta T$ for different accommodation coefficients $f_{ac}$ at $t=0.5\,\mathrm{ms}$. (b) Comparison with \eqref{eq:theta-app-05}, where $k_1=97.3$. The dashed line indicates equality between the abscissa and ordinate.}
	\label{fig:3-2-7}
\end{figure}

\subsubsection{Apparent contact angle}

Finally, we examine the effect of phase change on the apparent contact angle $\theta_a$, which provides a macroscopic measure of the phase-change-induced steepening of the dewetting rim, as observed in figure~\ref{fig:3-2-1}(a). Following the definition used in \S\ref{sec:3.1}, $\theta_a$ is defined as $2h_r/w$, as illustrated in figure~\ref{fig:2-1-1}(a). Figure~\ref{fig:3-2-7}(a) compares the apparent contact angles at $t=0.5\,\mathrm{ms}$ for different wall superheats $\Delta T$ and accommodation coefficients $f_{ac}$. As expected from the film-profile and interface-slope distributions discussed above, $\theta_a$ increases with both $\Delta T$ and $f_{ac}$. For example, for $\Delta T=16\,\mathrm{K}$ and $f_{ac}=0.4$, the apparent contact angle is increased from 0.06 to 0.093.

To describe the evaporation-induced increase in the apparent contact angle, we first recall the relation derived by \citet{hocking1995contact} from a lubrication-approximation-based analysis of an evaporating droplet,
\begin{equation} \label{eq:evap-h0}
	\theta_a^4 = \theta_e^4 + 12\,Ca_e\ln\left(\frac{e\theta_e l_R}{2\lambda_s\theta_a}\right).
\end{equation}
This relation describes the increase of the apparent contact angle with wall superheat. It should be noted, however, that in Hocking's derivation the contact line is assumed to remain static, with the evaporative mass loss compensated by liquid supply through the droplet base. In the present dewetting configuration, phase change and contact-line motion act simultaneously. Evaporation tends to increase $\theta_a$ relative to $\theta_e$, whereas the receding contact-line motion tends to reduce $\theta_a$, as described by \eqref{eq:app_inner} for isothermal dewetting. Motivated by these two limiting behaviours, we phenomenologically combine \eqref{eq:app_inner} and \eqref{eq:evap-h0} and propose the following relation for the apparent contact angle:
\begin{equation}\label{eq:theta-app-05}
	\theta_{a}^{3} = \left[\theta_{e}^{4} + 12\,Ca_e\ln\left(\frac{k_1\theta_el_R}{\lambda_s}\right)\right]^{3/4} - 9\,Ca \,\ln\!\left(\frac{2\,\theta_{e}\,w}{3\,\lambda_s}\right).
\end{equation}
Here, the first term on the right-hand side represents the cube of the evaporation-modified apparent contact angle in the absence of contact-line motion, while the second term accounts for the hydrodynamic correction associated with the receding contact line. The dimensionless constant $k_1$ is introduced as a fitting parameter in the logarithmic argument of the evaporation term to account for differences between the evaporating-droplet geometry considered by \citet{hocking1995contact} and the dewetting-rim geometry considered here, as well as the approximations involved in combining the two limiting relations.

Compared with the isothermal relation \eqref{eq:app_inner}, \eqref{eq:theta-app-05} may be viewed as replacing $\theta_e^3$ by the cube of an evaporation-modified apparent contact angle. It also has the same structure as the scale-separated relation for the apparent contact angle of \citet{zhang2022dewetting}. This agreement, however, holds only for this macroscopic quantity and does not contradict the results of \S\ref{sec:slope} on the local slope distribution. The numerical results shown in figure~\ref{fig:3-2-7}(a) are compared with \eqref{eq:theta-app-05} in figure~\ref{fig:3-2-7}(b), using a single value $k_1=97.3$ fitted to all the data. The data for different wall superheats and accommodation coefficients collapse reasonably well around the line of equality. This result indicates that \eqref{eq:theta-app-05} provides a reasonable description of the apparent contact angle of a dewetting film with phase change, at least for the small-equilibrium-contact-angle cases considered here.

\subsection{Comparison with lubrication equations}
\label{sec:3.3}

\begin{figure}[t]
	\centering
	\begin{subfigure}[t]{0.48\textwidth}
		\centering
		\includegraphics[width=\linewidth]{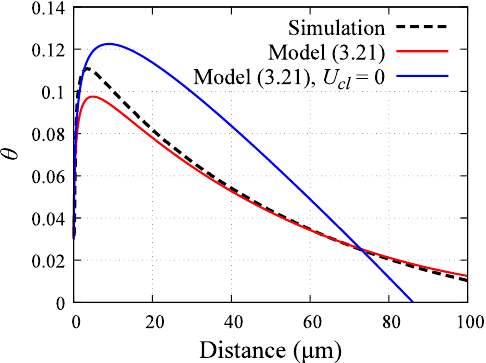}
		\subcaption{}
	\end{subfigure}
	\hfill
	\begin{subfigure}[t]{0.48\textwidth}
		\centering
		\includegraphics[width=\linewidth]{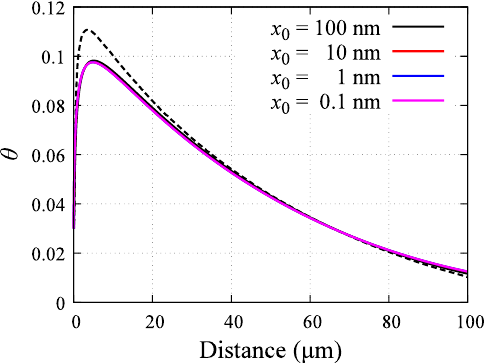}
		\subcaption{}
	\end{subfigure}
	\caption{Comparison of the DNS results with solutions of the lubrication model \eqref{eq:ivp} for $\Delta T=8\,\mathrm{K}$, $f_{ac}=0.4$ and $\theta_e=5^{\circ}$. (a) Interfacial slope distribution, including model solutions with the DNS contact-line speed and with $U_{cl}=0$. (b) Sensitivity of the model solution to the numerical integration starting point $x_0$, with the dashed curve denoting the corresponding DNS result.}
	\label{fig:3-3-2}
\end{figure}

In this section, we compare the DNS results with solutions obtained by solving the lubrication equation \eqref{lubrication} directly. In contrast to the asymptotic analysis in \S\ref{sec:slope}, the evaporation term is not treated here as a small perturbation of the Cox--Voinov solution. Returning \eqref{lubrication} to dimensional variables, we write it as the following first-order system, which is convenient for numerical integration with a Runge--Kutta method:
\begin{subequations}\label{eq:ivp}
%	\begin{equation}
%		\frac{\partial x}{\partial s} = \cos \theta
%	\end{equation}
	\begin{equation}\label{eq:hh}
		\frac{\partial h}{\partial x} = \theta,
	\end{equation}
	\begin{equation}\label{eq:theta}
		\frac{\partial\theta}{\partial x} = - \frac{1}{\sigma}\Delta p,
	\end{equation}
	%\begin{equation}\label{eq:dp}
	%	\frac{\partial\Delta p}{\partial s} =  - \frac{\mu_l}{h^2(h + 3 l_s)}\frac{Q}{L_e \rho_l} \frac{8\sin^3\theta}{2\theta \cos 2\theta - \sin 2\theta} - U_{cl} \frac{\mu_l}{h(h + 3 l_s)}\frac{4\sin^3 \theta}{2\theta - \sin 2\theta}
	%\end{equation}
	\begin{equation}\label{eq:dp}
		\frac{\partial\Delta p}{\partial x} = 3\,\mu_l \left[\frac{Q}{L_e \rho_l}\frac{1}{h^2(h + 3 \lambda_s)} -  \frac{U_{cl}}{h(h + 3  \lambda_s)}\right],
	\end{equation}
	%\begin{equation}\label{eq:dp2}
	%	\frac{\partial\Delta p}{\partial x} = 3\,\mu_l \left[\frac{Q}{L_v \rho_l}\frac{1}{(h + l_s)^3} -  \frac{U_{cl}}{(h + l_s)^2}\right]
	%\end{equation}
	%\begin{equation}
	%	\frac{\partial Q}{\partial s} = \frac{T_w - T_s(1 - \frac{\Delta p}{\rho_l L_e})}{{r\theta}/{\lambda_l} + R_{\Gamma}}
	%\end{equation}
	\begin{equation}\label{eq:Q}
		\frac{\partial Q}{\partial x} = \frac{T_w - T_s}{{h}/{k_l} + R_{\Gamma}}.
	\end{equation}
\end{subequations}
Here, $h$ and $\theta$ are the local film thickness and interfacial slope, respectively, $\Delta p=p_l-p_v$ is the pressure difference between the liquid and vapour phases, and $Q>0$ is the integrated heat flux measured from the contact line. Equation \eqref{eq:hh} is the small-slope geometric relation, while \eqref{eq:theta} follows from the normal-stress balance at the liquid--vapour interface. The two terms on the right-hand side of \eqref{eq:dp} represent two distinct contributions to the pressure gradient. The first is associated with evaporative mass loss and corresponds to the pressure gradient induced by a corner flow with a mass sink at the apex \citep{moffatt1980}. The second is associated with the receding motion of the contact line and corresponds to the pressure gradient induced by a wedge flow near a moving contact line \citep{moffatt1964}. Finally, \eqref{eq:Q} represents the one-dimensional heat-conduction approximation in the thin film, which has been verified against the DNS results in figure~\ref{fig:3-2-3}.

\begin{figure}[t]
	\centering
	\includegraphics[width=0.5\linewidth]{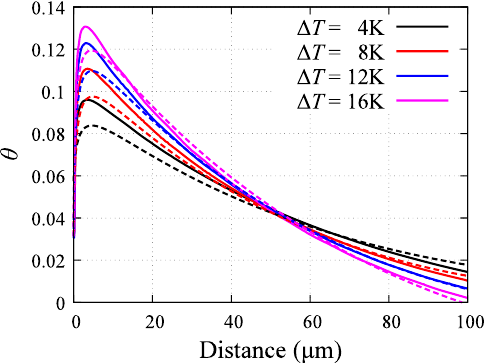}
	\caption{Comparison between interfacial-slope distributions from DNS (solid lines) and solutions of the lubrication model \eqref{eq:ivp} (dashed lines) for $\theta_e=5^{\circ}$ and $f_{ac}=0.4$ at different wall superheats.}
	\label{fig:3-3-3}
\end{figure}

Equation \eqref{eq:ivp} may also be regarded as the small-slope limiting form of the micro-region models for evaporating contact lines, originally proposed by \citet{mathieu2003} and subsequently adopted or extended by \citet{huber2017direct,zhang2022dewetting,torres2024coupling}. These models may include additional physical effects, such as the Kelvin effect, vapour recoil or Marangoni stresses. When the model is used to describe the micro-region itself, whose characteristic extent is typically on the micrometre scale, the hydrodynamic term associated with contact-line motion in \eqref{eq:dp} is often neglected. This simplification is based on the assumption that, within the micro-region, the slope modification is dominated by evaporation-induced flow. It also facilitates the usual shooting strategy, in which a vanishing-curvature condition is imposed at the outer edge of the micro-region, allowing the interfacial slope to approach a constant value. However, the evaporation term in \eqref{eq:dp} is not necessarily dominant as $h\to0$: with a finite interfacial heat-transfer resistance, $Q\propto h$ near the contact line, so that the ratio of the two terms tends approximately to $\varepsilon$ defined in \eqref{eq:ratio}, which is of order unity in the present cases (\S\ref{sec:slope}). Therefore, the hydrodynamic contribution is not negligible even within the micro-region. Moreover, in the present problem, we are interested not only in the micro-region but also in the overall interfacial-slope distribution resolved by the DNS. For this purpose, the contact-line-motion term must be retained, as demonstrated below.

%Model 1: corner flow with a mass sink at the apex. Then, the pressure gradient along the interface is \citep{moffatt1980}
%\begin{equation}
%	\frac{\partial \Delta p_1}{\partial r} =  \frac{\partial p_l}{\partial r} = - \frac{\mu_l}{r^3}\frac{Q}{L_v \rho_l} \frac{8}{2\theta \cos 2\theta - \sin 2\theta}
%\end{equation}
%where $Q > 0$ is integrated heat flux.
%
%Model 2: wedge flow near the moving contact line. Then, the pressure gradient along the interface is \citep{moffatt1964}
%\begin{equation}
%	\frac{\partial \Delta p_2}{\partial r} = \frac{\partial p_l}{\partial r} = - U_{cl} \frac{\mu_l}{r^2}\frac{4\sin \theta}{2\theta - \sin 2\theta}
%\end{equation}
%where $U_{cl} > 0$ is the contact line speed.

%From the Gibbs-Thompson relation, the interface saturation temperature is related to the pressure difference across the interface \citep{stephan1994}
%\begin{equation}
%	T_{i} = T_{\mathrm{sat}}(p_v)\cdot\left(1 - \frac{\Delta p}{L_v \rho_l}\right)
%\end{equation}

%Then we can estimate the heat flux q across the liquid film,
%\begin{equation}
%	q = \frac{T_w - T_{i}}{{h}/{\lambda_l} + R_{i}}
%\end{equation}
%where $T_w$ is the solid wall temperature and interfacial thermal resistance $R_i$ is defined as
%\begin{equation}
%	R_{\Gamma} = \frac{2-f}{2f}\frac{T_s\sqrt{2\pi R_g T_s}(\rho_l - \rho_v)}{\rho_l \rho_v L_v^2} \approx \frac{2-f}{2f}\frac{T_s\sqrt{2\pi R_g T_s}}{\rho_v L_v^2}
%\end{equation}
%where $f$ is the accommodation coefficient.

To integrate \eqref{eq:ivp}, the initial conditions also need to be specified. Since the system is singular at the contact line, where $h=0$, the integration is started at a small distance $x_0>0$ from the contact line. The initial conditions are prescribed as
\begin{subequations} \label{eq:initialvalue}
%\begin{equation}
%	\left. x\right|_{s=0} = 0
%\end{equation}
%\begin{equation}
%	\left. h\right|_{x=x_0} = \theta_{\mathrm{e}} x_0
%\end{equation}
%\begin{equation}
%	\left. \theta\right|_{x=x_0} = \theta_{\mathrm{e}}
%\end{equation}
%\begin{equation}
%	\left. \Delta p\right|_{x=x_0} = \Delta p_{\mathrm{shoot}}
%\end{equation}
%\begin{equation}
%	\left. Q\right|_{x=x_0} = \frac{T_w - T_s}{{\theta_{\mathrm{e}} x_0}/{\lambda_l} + R_{\Gamma}}x_0
%\end{equation}

%\begin{equation}
%	\left. x\right|_{x=0} = x_0
%\end{equation}
\begin{equation}
	\left. h\right|_{x=x_0} = \theta_0 x_0,
\end{equation}
\begin{equation}
	\left. \theta\right|_{x=x_0} = \theta_0,
\end{equation}
\begin{equation}
	\left. \Delta p\right|_{x=x_0} = \Delta p_{\mathrm{shoot}},
\end{equation}
\begin{equation}\label{eq:Q0}
	\left. Q\right|_{x=x_0} = \frac{k_l\Delta T}{ \theta_0}\ln\left(\frac{l_R + \theta_0x_0}{l_R}\right).
\end{equation}
\end{subequations}
Here, the interface in the unresolved region $0<x<x_0$ is approximated by a straight wedge of slope $\theta_0$. This slope is not necessarily equal to the imposed equilibrium contact angle $\theta_e$, since the interfacial slope resolved by the DNS near the contact line can differ from this prescribed value, as shown in figure~\ref{fig:3-2-4}. In practice, $\theta_0$ is taken as the first interfacial-slope value resolved by the DNS outside the contact-line cell. The pressure difference $\Delta p_{\mathrm{shoot}}$ is treated as a shooting parameter. Equation \eqref{eq:Q0} is obtained by integrating \eqref{eq:Q} over $0<x<x_0$ under the straight-wedge approximation, using $h=\theta_0 x$.

\begin{figure}[t]
	\centering
	\includegraphics[width=0.5\linewidth]{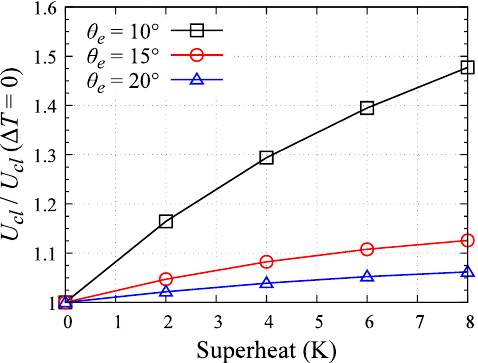}
	\caption{Dependence of the normalized contact-line speed on wall superheat at $t=0.4\,\mathrm{ms}$ and $f_{ac}=0.4$ for different equilibrium contact angles.}
	\label{fig:3-4-1}
\end{figure}

The numerical solution of \eqref{eq:ivp} for a representative case with $\theta_e=5^{\circ}$, $f_{ac}=0.4$ and $\Delta T=8\,\mathrm{K}$ is shown in figure~\ref{fig:3-3-2}(a). The slip length $\lambda_s$ in \eqref{eq:dp} is set to the same value as that used in the DNS. The contact-line speed $U_{cl}$ in \eqref{eq:dp} and the initial slope $\theta_0$ in \eqref{eq:initialvalue} are obtained from the corresponding DNS result. The initial pressure difference $\Delta p_{\mathrm{shoot}}$ is then determined by minimizing the root-mean-square difference between the interfacial slopes predicted by the lubrication model and those obtained from the DNS:
\begin{equation}
	\mathcal{E}(\Delta p_{\mathrm{shoot}})
	=
	\left[
	\frac{1}{N}\sum_{i=1}^{N}
	\left(
	\theta_{\mathrm{model}}(x_i;\Delta p_{\mathrm{shoot}})
	-
	\theta_{\mathrm{DNS}}(x_i)
	\right)^2
	\right]^{1/2},
	\label{eq:shooting_error}
\end{equation}
where $N$ is the number of DNS data points in the fitting interval $0<x<100\,\mathrm{\mu m}$. Overall, the lubrication model reproduces the DNS result reasonably well, although it underpredicts the maximum interfacial slope in the transition region in which both evaporation-induced flow and the hydrodynamic flow associated with contact-line recession are important. To further clarify the role of contact-line motion, figure~\ref{fig:3-3-2}(a) also shows the solution of \eqref{eq:ivp} obtained by setting $U_{cl}=0$, with $\Delta p_{\mathrm{shoot}}$ again determined by minimizing \eqref{eq:shooting_error}. This solution captures the initial increase in interfacial slope near the contact line but fails to reproduce its subsequent variation. In particular, beyond the slope-increasing region, the predicted slope profile bends in a direction opposite to that observed in the DNS. Figure~\ref{fig:3-3-2}(b) shows the sensitivity of the numerical solution to the integration starting point $x_0$. Values of $x_0$ ranging from $0.1\,\mathrm{nm}$ to $100\,\mathrm{nm}$ are considered, with the largest value being comparable to the grid spacing used in the DNS. The resulting profiles are almost indistinguishable, indicating that the model solution is insensitive to the precise choice of $x_0$ over this range.
%In this localized region, the leading-order lubrication model may not fully capture the detailed coupling among phase change, interfacial curvature and liquid flow. 

Figure~\ref{fig:3-3-3} compares the solutions of \eqref{eq:ivp} with the DNS results obtained at different wall superheats. For each wall superheat, $U_{cl}$, $\theta_0$ and $\Delta p_{\mathrm{shoot}}$ are determined in the same manner as in figure~\ref{fig:3-3-2}. Similar to the results shown in figure~\ref{fig:3-3-2}(a), the lubrication model reproduces the overall interfacial-slope distribution reasonably well for all wall superheats considered. The main discrepancy remains localized near the maximum of the interfacial slope, where the model tends to underestimate the peak value. Nevertheless, the difference between the DNS and model predictions decreases as $\Delta T$ increases. Together with the comparison obtained by setting $U_{cl}=0$ in figure~\ref{fig:3-3-2}(a), these results confirm that retaining the contact-line-motion term is essential for describing the overall interfacial-slope distribution.

\subsection{Effects of contact angle}
\label{sec:3.4}

\begin{figure}[t]
	\centering
	\begin{subfigure}[t]{0.48\textwidth}
		\centering
		\includegraphics[width=\linewidth]{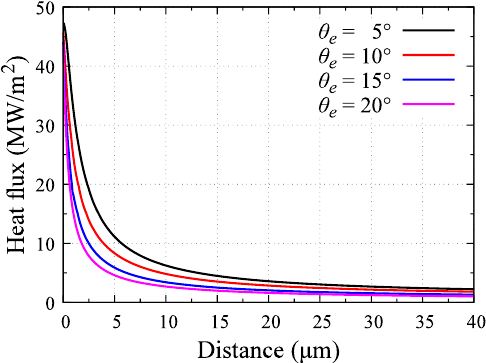}
		\subcaption{}
	\end{subfigure}
	\hfill
	\begin{subfigure}[t]{0.48\textwidth}
		\centering
		\includegraphics[width=\linewidth]{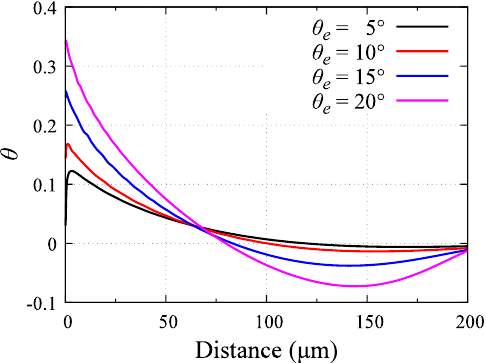}
		\subcaption{}
	\end{subfigure}
	\caption{Interfacial heat flux (a) and interfacial slope (b) as functions of the distance from the contact line for different equilibrium contact angles $\theta_e$. Results are shown at $t=0.4\,\mathrm{ms}$ for $f_{ac}=0.4$ and $\Delta T=12\,\mathrm{K}$.}
	\label{fig:3-4-4}
\end{figure}

The preceding analysis has focused mainly on the case $\theta_e=5^{\circ}$, for which the capillary-driven retraction is weak and the influence of phase change is therefore most clearly visible. We now examine how this influence varies with the equilibrium contact angle. Figure~\ref{fig:3-1-1} compares the film profiles obtained for different $\theta_e$ with and without phase change. In the presence of phase change, the contact line recedes farther and the rim becomes steeper, indicating an increase in the apparent contact angle. This evaporation-induced modification is most pronounced for $\theta_e=5^{\circ}$ and becomes progressively less significant as $\theta_e$ increases. The same trend is quantified in figure~\ref{fig:3-4-1}, where the contact-line speed, normalized by its isothermal value, is plotted as a function of wall superheat. For all equilibrium contact angles considered, the normalized contact-line speed increases with $\Delta T$, but the relative increase is strongly reduced at larger $\theta_e$. In particular, even for $\theta_e=10^{\circ}$, the acceleration induced by phase change is much weaker than that observed for $\theta_e=5^{\circ}$ in figure~\ref{fig:3-2-2}(b). This behaviour can be understood by noting that the capillary driving force, and hence the isothermal dewetting speed, increases rapidly with $\theta_e$, as shown in figure~\ref{fig:3-1-2}. As the capillary-induced contact-line motion becomes stronger, the additional acceleration caused by phase change becomes less significant relative to the isothermal dewetting speed.

\begin{figure}[t]
	\centering
	\begin{subfigure}[t]{0.5\textwidth}
		\centering
		\includegraphics[width=\linewidth]{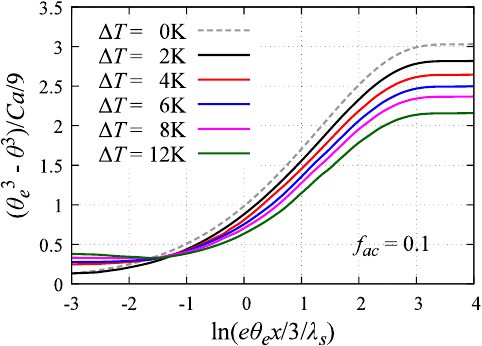}
		\subcaption{}
	\end{subfigure}
	\hfill
	\begin{subfigure}[t]{0.46\textwidth}
		\centering
		\includegraphics[width=\linewidth]{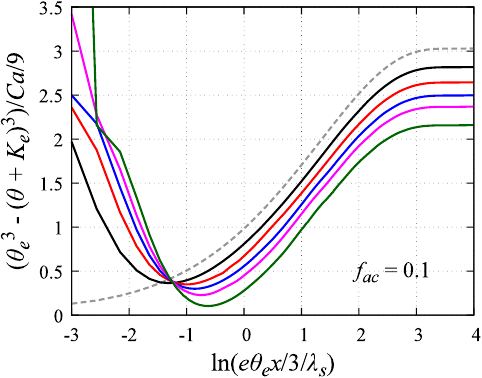}
		\subcaption{}
	\end{subfigure}
	\begin{subfigure}[t]{0.5\textwidth}
		\centering
		\includegraphics[width=\linewidth]{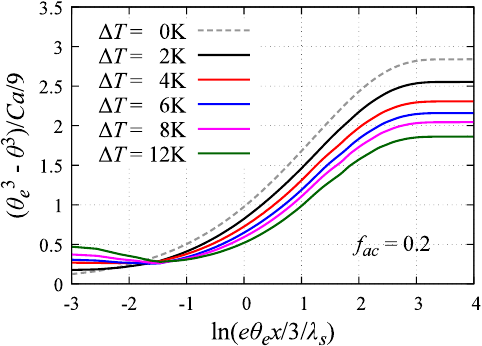}
		\subcaption{}
	\end{subfigure}
	\hfill
	\begin{subfigure}[t]{0.46\textwidth}
		\centering
		\includegraphics[width=\linewidth]{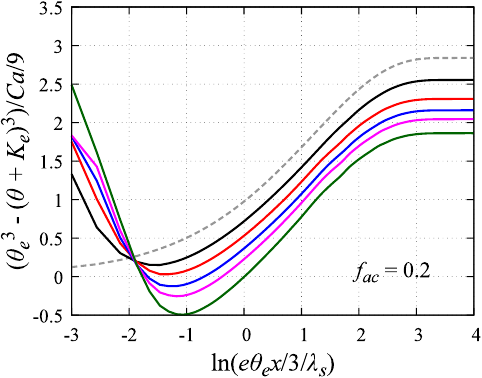}
		\subcaption{}
	\end{subfigure}
	\caption{Rescaled interfacial-slope distributions for $\theta_e=10^{\circ}$, taken after the slope profiles have reached a quasi-steady state. Results are shown for $f_{ac}=0.1$ in (a,b) and $f_{ac}=0.2$ in (c,d), with different curves corresponding to different wall superheats. Panels (a,c) use the Cox--Voinov scaling implied by \eqref{eq:zhang}, whereas panels (b,d) use the evaporation-corrected form \eqref{solution-large-x}, with $K_e = Ca_e l_R \left[1.5\ln\left(\theta_e x/l_R\right)+2.43\right]/\left(\theta_e^4 x\right)$ included in the ordinate. The dashed curves denote the isothermal-dewetting result at $\theta_e=10^{\circ}$.}
	\label{fig:3-4-5}
\end{figure}

Figure~\ref{fig:3-4-4} compares the interfacial heat-flux and slope distributions as functions of the distance from the instantaneous contact line for different equilibrium contact angles. As $\theta_e$ increases, the heat flux decays more rapidly away from the contact line. This behaviour is expected because, at a given distance from the contact line, a larger equilibrium contact angle corresponds to a thicker liquid film, approximately $h\sim \theta_e x$, and therefore to a smaller conductive heat flux through the film. Consequently, the region over which evaporation can significantly modify the interfacial shape becomes increasingly confined near the contact line. This trend is reflected in the interfacial-slope distributions shown in figure~\ref{fig:3-4-4}(b). For small $\theta_e$, especially $\theta_e=5^{\circ}$, a clear phase-change-induced increase in the interfacial slope is observed near the contact line. By contrast, this increasing-slope region becomes much less pronounced as $\theta_e$ increases, and is barely visible for $\theta_e=15^{\circ}$ and $20^{\circ}$, even for the relatively large accommodation coefficient and wall superheat considered in figure~\ref{fig:3-4-4}. 
At larger $\theta_e$, both the more localized heat flux and the faster contact-line recession (figure~\ref{fig:3-1-2}) reduce $\varepsilon=Ca_e/(\theta_e Ca)$ defined in \eqref{eq:ratio}, which decreases from about 4 at $\theta_e=5^\circ$ to about 0.2 at $\theta_e=20^\circ$, so that the evaporative contribution becomes subdominant throughout the near-contact-line region at large $\theta_e$.

\begin{figure}[t]
	\centering
	\includegraphics[width=0.7\linewidth]{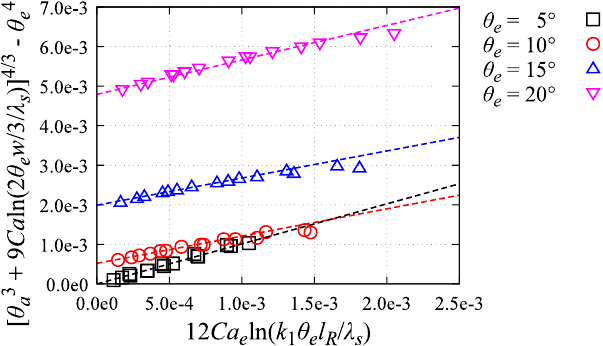}
	\caption{Comparison of the apparent contact angle $\theta_a=2h_r/w$ with \eqref{eq:theta-app-05} using $k_1=97.3$ for different $\theta_e$. The data include cases with $f_{ac}=0.1$, $0.2$ and $0.4$. The black dashed line indicates equality between the abscissa and ordinate, while the other dashed straight lines serve as guides to the eye.}	
	\label{fig:3-4-3}
\end{figure}

Figure~\ref{fig:3-4-5} compares the interfacial-slope distributions for $\theta_e=10^{\circ}$ with the Cox--Voinov-type relation \eqref{eq:zhang} and the evaporation-corrected asymptotic expression \eqref{solution-large-x}. Results for larger equilibrium contact angles are not shown, because the phase-change-induced modification of the slope distribution is already very weak for those cases, as indicated in figure~\ref{fig:3-4-4}(b). Panels (a,c) show the comparison with the Cox--Voinov-type relation \eqref{eq:zhang}. In this logarithmic representation, the slopes of the curves in the intermediate region decrease slightly as the wall superheat increases, indicating that the Cox--Voinov relation alone does not fully collapse the data. This behaviour is consistent with that observed for $\theta_e=5^{\circ}$ in figure~\ref{fig:3-2-5}(b). When the evaporation correction in \eqref{solution-large-x} is included, as shown in panels (b,d), the curves obtained at different wall superheats become nearly parallel to the corresponding isothermal result in the intermediate region. The same integration constant as that used for $\theta_e=5^{\circ}$ in figure~\ref{fig:3-2-6}, namely $C_1=-2.43$, is adopted here without any additional adjustment. This agreement indicates that \eqref{solution-large-x} remains applicable for $\theta_e=10^{\circ}$ with $f_{ac}=0.1$ and $0.2$, for which the influence of phase change on the slope distribution is relatively weak.

Finally, we examine the applicability of \eqref{eq:theta-app-05} for predicting the apparent contact angle at different equilibrium contact angles. The results are shown in figure~\ref{fig:3-4-3}, where the same value $k_1=97.3$ as that used in figure~\ref{fig:3-2-7}(b) is adopted. For $\theta_e=5^{\circ}$, the data collapse well onto the line of equality, as shown previously in figure~\ref{fig:3-2-7}(b). For larger equilibrium contact angles, however, the data obtained at different accommodation coefficients and wall superheats no longer collapse onto this line, although they still approximately follow a linear trend. These approximate linear trends indicate that \eqref{eq:theta-app-05} still provides a useful scaling representation of the data, although the quantitative agreement with this relation deteriorates at larger $\theta_e$ when the same value of $k_1$ is used. The deviation may indicate that a single value of $k_1$ does not fully capture the dependence of the evaporation contribution on $\theta_e$, or that the hydrodynamic correction based on the rim half-width $w$ becomes less accurate at larger $\theta_e$.

\subsection{Effects of slip length}
\label{sec:3.5}

%\begin{figure}[t]
%	\centering
%	\begin{subfigure}[t]{0.47\textwidth}
%		\centering
%		\includegraphics[width=\linewidth]{figs/fig3-5-1a.pdf}
%		\subcaption{}
%	\end{subfigure}
%	\hfill
%	\begin{subfigure}[t]{0.496\textwidth}
%		\centering
%		\includegraphics[width=\linewidth]{figs/fig3-5-1b.pdf}
%		\subcaption{}
%	\end{subfigure}
%	\begin{subfigure}[t]{0.47\textwidth}
%		\centering
%		\includegraphics[width=\linewidth]{figs/fig3-5-1c.pdf}
%		\subcaption{}
%	\end{subfigure}
%	\hfill
%	\begin{subfigure}[t]{0.495\textwidth}
%		\centering
%		\includegraphics[width=\linewidth]{figs/fig3-5-1d.pdf}
%		\subcaption{}
%	\end{subfigure}
%	\caption{Interface slope distribution at $t=0.4\,\mathrm{ms}$. $\theta=5^{\circ}$.  $f=0.1$ in (a) and (b). $f=0.4$ in (c) and (d).}
%	\label{fig:3-5-1}
%\end{figure}

In this section, we assess the influence of the slip length on the DNS results. In the preceding sections, the slip length was fixed at $\lambda_s=500\,\mathrm{nm}$, which is much larger than a physical slip length, typically of molecular scale. This artificially large value was adopted as a computational compromise, because the grid spacing must be smaller than the prescribed slip length in order to resolve the slip region. To examine the sensitivity of the results to this choice, we repeat selected simulations with the slip length reduced by one order of magnitude, namely $\lambda_s=50\,\mathrm{nm}$. The grid is refined accordingly so that the reduced slip length remains resolved, with $\lambda_s>3\Delta$. We focus on the interfacial-slope distribution near the contact line, which is expected to be particularly sensitive to the slip length and the local grid resolution.

\begin{figure}[t]
	\centering
	\begin{subfigure}[t]{0.48\textwidth}
		\centering
		\includegraphics[width=\linewidth]{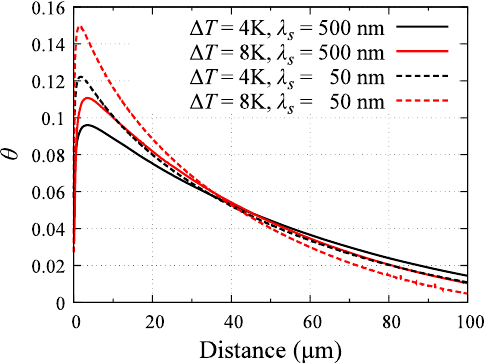}
		\subcaption{}
	\end{subfigure}
	\hfill
	\begin{subfigure}[t]{0.48\textwidth}
		\centering
		\includegraphics[width=\linewidth]{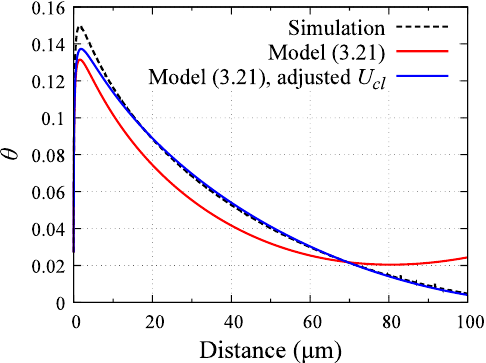}
		\subcaption{}
	\end{subfigure}
	\caption{Effect of slip length on the interfacial-slope distribution for $\theta_e=5^{\circ}$ and $f_{ac}=0.4$. (a) Simulation results obtained with $\lambda_s=500\,\mathrm{nm}$ and $50\,\mathrm{nm}$ at $\Delta T=4\,\mathrm{K}$ and $8\,\mathrm{K}$. (b) Comparison between the DNS result and solutions of the lubrication model \eqref{eq:ivp} for $\lambda_s=50\,\mathrm{nm}$ and $\Delta T=8\,\mathrm{K}$. The model solutions are obtained with $U_{cl}=U_{cl}^{\mathrm{DNS}}$ and $U_{cl}=0.86U_{cl}^{\mathrm{DNS}}$, respectively.}
	\label{fig:3-5-2}
\end{figure}

The results are shown in figure~\ref{fig:3-5-2}(a), where two different wall superheats are considered for $\theta_e=5^{\circ}$ and $f_{ac}=0.4$. Reducing the slip length from $\lambda_s=500\,\mathrm{nm}$ to $50\,\mathrm{nm}$ leads to a more pronounced increase in the interfacial slope close to the contact line. This behaviour is consistent with the lubrication model: according to \eqref{eq:dp}, reducing $\lambda_s$ enhances both pressure-gradient contributions through the denominators containing $h+3\lambda_s$. In the slope-increasing region, where the evaporation-related pressure gradient term exceeds the opposing hydrodynamic term, reducing $\lambda_s$ further enhances the net evaporation-driven bending of the interface. In figure~\ref{fig:3-5-2}(b), the DNS result for $\lambda_s=50\,\mathrm{nm}$ and $\Delta T=8\,\mathrm{K}$ is compared with solutions of the lubrication model \eqref{eq:ivp}. When the contact-line speed measured in the DNS is used, the model captures the overall trend of the interfacial-slope distribution, although quantitative deviations remain. A closer agreement is obtained when the contact-line speed used in the model is reduced to $0.86U_{cl}^{\mathrm{DNS}}$. This sensitivity to $U_{cl}$ indicates that the hydrodynamic contribution in \eqref{eq:dp} plays an important role in shaping the overall interfacial-slope distribution.

\begin{figure}[t]
	\centering
	\includegraphics[width=0.9\linewidth]{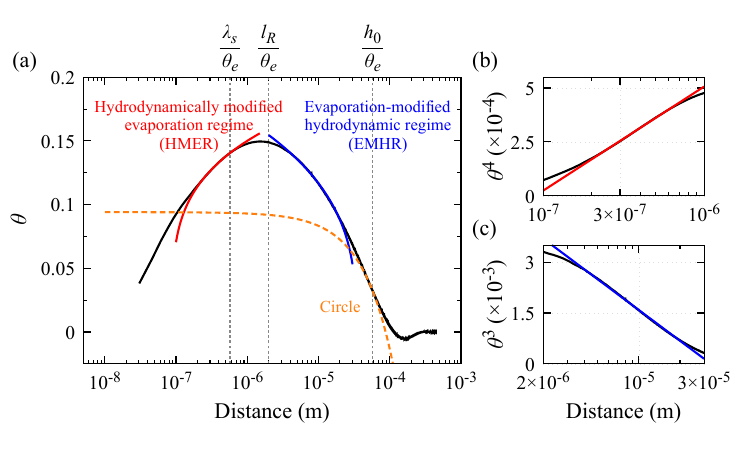}
	\caption{Interfacial slope versus the distance from the contact line for $\theta_e=5^{\circ}$, $f_{ac}=0.4$, $\Delta T=8\,\mathrm{K}$ and $\lambda_s=50\,\mathrm{nm}$. The red and blue curves denote linear fits of $\theta^4$ and $\theta^3$ against $\ln x$ in the hydrodynamically modified evaporation regime (HMER) and the evaporation-modified hydrodynamic regime (EMHR), respectively (fitting intervals given in Appendix~\ref{app:local_log_slope}). Panels ({b}) and ({c}) show these fits in the $\theta^4$ and $\theta^3$ representations, with fitted logarithmic slopes of $2.09\times10^{-4}$ and $-1.31\times10^{-3}$, respectively. The vertical dashed lines in panel ({a}) indicate three distances, $\lambda_s/\theta_e$, $l_R/\theta_e$ and $h_0/\theta_e$.}
	\label{fig:3-5-3}
\end{figure}

In \S\ref{sec:slope}, we showed that, outside the micro-region, the effect of phase change cannot be captured solely by adjusting the Voinov angle and Voinov length in the Cox--Voinov-type relation \eqref{eq:zhang}. Here, we use the reduced-slip-length case to examine the complementary question of whether contact-line motion can be neglected within the micro-region itself. This case is better suited to this purpose because the slope-increasing region near the contact line is more pronounced. For a stationary contact line, \citet{hocking1995contact} obtained the following logarithmic relation for the interfacial slope in the intermediate region:
\begin{equation}
	\theta^4 = \theta_e^4 + 12\,Ca_e\ln\left(\frac{e\theta_e x}{3\lambda_s}\right).
	\label{eq:hocking-var}
\end{equation}
This relation predicts a logarithmic coefficient of $12Ca_e$ for $\theta^4$. Figure~\ref{fig:3-5-3}(a) shows a typical interfacial-slope distribution for $\lambda_s=50\,\mathrm{nm}$, with three characteristic horizontal distances, $\lambda_s/\theta_e$, $l_R/\theta_e$ and $h_0/\theta_e$, indicated for reference. At $x<l_R/\theta_e$, the slope increases with distance, and panel ({b}) shows an approximately linear variation of $\theta^4$ with $\ln x$. The fitted logarithmic coefficient is $2.09\times10^{-4}$, approximately 25\% of the stationary-contact-line prediction $12Ca_e$. As shown by the approximate local slope relation \eqref{eq:app_B4_local} derived from the lubrication equation in Appendix~\ref{app:local_log_slope}, contact-line recession counteracts evaporation-induced interfacial bending, while finite slip and the local interface geometry also affect the logarithmic coefficient. The corresponding estimate of $2.23\times10^{-4}$ agrees reasonably well with the DNS-fitted value. We therefore refer to this slope-increasing region as the hydrodynamically modified evaporation regime (HMER).

Farther from the contact line ($x>l_R/\theta_e$), the interfacial slope decreases with distance, and figure~\ref{fig:3-5-3}(c) shows an approximately linear variation of $\theta^3$ with $\ln x$. This behaviour resembles the Cox--Voinov variation \eqref{eq:zhang} associated with receding contact-line motion, but the fitted logarithmic coefficient is $-1.31\times10^{-3}$, approximately 58\% of the isothermal Cox--Voinov prediction $-9Ca$. As shown by the approximate local slope relation \eqref{eq:app_B3_local} derived in Appendix~\ref{app:local_log_slope}, evaporation introduces a positive correction to the negative logarithmic coefficient associated with contact-line recession. However, the corresponding estimate of $-7.48\times10^{-4}$ underestimates the magnitude of the DNS-fitted coefficient. We therefore refer to this slope-decreasing region as the evaporation-modified hydrodynamic regime (EMHR). At larger distances, with $x\sim h_0/\theta_e$, the interfacial-slope distribution can be approximated by that of a circular arc, as shown in figure~\ref{fig:3-5-3}(a).

\section{Conclusions}
\label{sec:con}

We have performed two-dimensional direct numerical simulations of micrometre-scale film dewetting with phase change on a superheated wall, in which a finite interfacial heat-transfer resistance and Navier slip regularize the thermal and viscous singularities at the contact line. Contact-line-specific numerical treatments, namely a modified ghost fluid method and a local evaluation of the liquid-side heat flux, enable grid-converged results for equilibrium contact angles down to $5^{\circ}$. In the isothermal limit, the simulated contact-line speed and apparent contact angle agree reasonably well with the asymptotic predictions of \citet{snoeijer2010asymptotic}. The slope expression derived from this theory agrees better with the DNS than its linearised Cox--Voinov form, particularly at small contact angles.

Phase change accelerates contact-line recession and increases the apparent contact angle, both effects strengthening with increasing wall superheat and accommodation coefficient. For $\theta_e=5^{\circ}$ and $f_{ac}=0.4$, the contact-line speed at $\Delta T=8\,\mathrm{K}$ exceeds its isothermal value by more than an order of magnitude, and the apparent contact angle rises from 0.06 to 0.093 at $\Delta T=16\,\mathrm{K}$. The interfacial heat flux is localized near the contact line and follows a local one-dimensional conduction model. The resulting evaporation drives liquid towards the contact line, whereas capillary-driven retraction dominates farther away, producing a near-contact-line region in which the interfacial slope increases with distance and contributes to the larger apparent contact angle. The influence of evaporation weakens rapidly as $\theta_e$ increases, because faster capillary dewetting and thicker films at a given distance from the contact line reduce the relative contribution of evaporation.

For the micrometre-scale films investigated here, the influence of phase change is not confined to an inner micro-region but extends into the outer region. The rescaled slope distributions do not collapse onto a Cox--Voinov form, and their logarithmic slopes decrease with wall superheat, which cannot be accounted for by adjusting the Voinov angle and length alone. A scaling argument based on the lubrication equation explains this failure: the ratio of evaporative to hydrodynamic liquid flux, $R(\hat{h})\simeq\varepsilon\ln(1+\hat{h})/\hat{h}$ with $\varepsilon=Ca_e/(\theta_e Ca)$, 
shows that a Cox--Voinov region can develop outside the micro-region only if $R(h_0/l_R)\ll1$, which is possible only in films much thicker than $l_R$, whereas neglecting contact-line motion within the micro-region further requires $\varepsilon\gg1$. For $\theta_e=5^\circ$, $h_0/l_R\approx3$--$30$, giving $R(h_0/l_R)\approx0.2$--$0.5$, and $\varepsilon\approx0.5$--$4$, so neither condition is met. Thus, evaporation-induced flow and contact-line motion act over overlapping length scales and must be treated together. Our first-order evaporation correction to the Cox--Voinov solution approximately recovers the isothermal logarithmic slope in the weak-phase-change regime, and a lubrication model reproduces the overall slope distribution only when both contributions are retained. A phenomenological relation combining the prediction of \citet{hocking1995contact} for evaporating static contact lines with the Cox--Voinov-type relation of \citet{snoeijer2010asymptotic} for isothermal dewetting collapses the apparent contact angles at $\theta_e=5^{\circ}$ with a single fitted constant,  although its quantitative agreement deteriorates at larger $\theta_e$.

At a reduced slip length, $\lambda_s=50\,\mathrm{nm}$, the slope distribution exhibits two regions of approximately logarithmic variation, separated at $x\approx l_R/\theta_e$. In the hydrodynamically modified evaporation regime (HMER), closer to the contact line, $\theta^4$ varies linearly with $\ln x$, as in the stationary-contact-line solution of \citet{hocking1995contact}, but with a coefficient of only about 25\,\% of $12\,Ca_e$, reduced by contact-line recession and finite slip. In the evaporation-modified hydrodynamic regime (EMHR), farther away, $\theta^3$ varies linearly with $\ln x$, as in the Cox--Voinov law, but with a coefficient of about 58\,\% of $-9\,Ca$, reduced in magnitude by evaporation. Together with $\varepsilon=O(1)$, these results indicate that contact-line motion cannot be neglected even within the micro-region.

%The present simulations use a slip length much larger than molecular scales and neglect the Kelvin effect, vapour recoil and Marangoni stresses. These effects are small under the present conditions but may become important as the contact-line region is resolved towards physically realistic slip lengths. Extending the DNS to such scales would allow a more direct comparison with experiments on evaporating films and microlayers.

The present simulations use slip lengths much larger than molecular scales and neglect the Kelvin effect, vapour recoil and Marangoni stresses. The first two effects are estimated to be small under the present conditions (\S\ref{sec:2.1}), but may become important when the contact-line region is resolved down to physically realistic slip lengths. Such slip lengths could be approached either by refining the grid further at fixed film thickness or by reducing the film thickness to a few hundred nanometres while keeping $h_0/\lambda_s$ fixed. In the latter case, however, the film thickness would be comparable to or smaller than $l_R$, which may correspond to a different regime. In both cases, the nanometre-scale grid spacing would severely restrict the time step when surface tension is treated explicitly, motivating the use of implicit schemes \citep{popinet2018numerical}. Finally, the lubrication model \eqref{eq:ivp}, which retains contact-line motion and captures the overall simulated slope distributions, could provide a basis for developing subgrid models of the contact-line region in multiscale simulations of nucleate boiling.

\section*{Funding}
This work has received funding from the European Research Council (ERC) under the European Union’s Horizon 2020 research and innovation programme (project TRUFLOW, grant agreement number 883849). We are grateful to GENCI for generous allocation on Irene supercomputers (grant agreement numbers A0212B14629 and A0202B17546). This work has been supported by the National Natural Science Foundation of China (Grants No.~12302343 and No.~12372266), and by the Postdoctoral Science Foundation of China (2023M733461).

\section*{Declaration of interests}
The authors report no conflict of interest.

% \backsection[Supplementary data]{\label{SupMat}Supplementary material and movies are available at \\https://doi.org/10.1017/jfm.2019...}
%
% \backsection[Acknowledgements]{Acknowledgements may be included at the end of the paper, before the References section or any appendices. Several anonymous individuals are thanked for contributions to these instructions.}
%
% \backsection[Funding]{Please provide details of the sources of financial support for all authors, including grant numbers. Where no specific funding has been provided for research, please provide the following statement: "This research received no specific grant from any funding agency, commercial or not-for-profit sectors." }
%
% \backsection[Declaration of interests]{A Competing Interests statement is now mandatory in the manuscript PDF. Please note that if there are no conflicts of interest, the declaration in your PDF should read as follows: {\bf Declaration of Interests}. The authors report no conflict of interest.}
%
% \backsection[Data availability statement]{The data that support the findings of this study are openly available in [repository name] at http://doi.org/[doi], reference number [reference number]. See JFM's \href{https://www.cambridge.org/core/journals/journal-of-fluid-mechanics/information/journal-policies/research-transparency}{research transparency policy} for more information}
%
% \backsection[Author ORCIDs]{Authors may include the ORCID identifers as follows.  F. Smith, https://orcid.org/0000-0001-2345-6789; B. Jones, https://orcid.org/0000-0009-8765-4321}
%
% \backsection[Author contributions]{Authors may include details of the contributions made by each author to the manuscript'}

%\appendix
\begin{appen}
	
\renewcommand{\theHsection}{\Alph{section}}
\renewcommand{\theHequation}{\Alph{section}.\arabic{equation}}

\section{Asymptotic rim solution near the contact line}\label{appB}

In this appendix, we briefly summarise the asymptotic analysis of \citet{snoeijer2010asymptotic} for the rim profile of a dewetting film and then derive the logarithmic variation of the interface slope in the near-contact-line region used in \S\ref{sec:3.1}. We consider the film profile in a frame moving with the contact line. Under the lubrication approximation, the governing equation reads
\begin{equation}\label{eq:govern}
	\partial_t h - U_{cl}\,\partial_x h
	+ \frac{\sigma}{3\mu}\,\partial_x\!\left[h^2\,(h+3\lambda_s)\,\partial_{xxx}h\right]=0,
\end{equation}
where $h$ is the film thickness and $U_{cl}$ is the contact-line speed.

For a quasi-steady rim profile, the unsteady term in \eqref{eq:govern} is neglected. Integrating once with respect to $x$, and setting the integration constant to zero using $h=0$ at the contact line, we obtain
\begin{equation}\label{eq:simply}
	h_{xxx} = \frac{3\,Ca}{h\,(h+3\lambda_s)},
\end{equation}
where $Ca=\mu U_{cl}/\sigma$ is the capillary number.

We introduce the rescaled variables
\begin{equation}
	h(x)=3\lambda_s\,H(\xi), 
	\quad 
	\xi=\frac{x\,\theta_e}{3\lambda_s}.
\end{equation}
Equation \eqref{eq:simply} then becomes
\begin{equation}\label{eq:H}
	H^{\prime\prime\prime}=\frac{\delta}{H(H+1)},
\end{equation}
where primes denote differentiation with respect to $\xi$, and $\delta = {3\,Ca}/{\theta_e^{3}}$ is the rescaled capillary number. In the present analysis, we assume $\delta\ll 1$.

We focus on the region adjacent to the contact line but outside the slip region, where the film thickness is large compared with the slip length, i.e. $h\gg 3\lambda_s$ or equivalently $H\gg 1$. In this region, \eqref{eq:H} reduces to
\begin{equation}
	H^{\prime\prime\prime}\approx\frac{\delta}{H^{2}}.
\end{equation}
Introducing $H(\xi)=\delta^{1/3}\,y(\xi)$, we obtain
\begin{equation}\label{eq:y}
	y^{\prime\prime\prime}=\frac{1}{y^{2}}.
\end{equation}

A parametric solution of \eqref{eq:y} satisfying $y(0)=0$ can be written as
\begin{equation}\label{eq:paraSol}
	\begin{aligned}
		\xi &= \frac{2^{1/3}\pi\,\mathrm{Ai}(s)}
		{\beta\bigl[\alpha\,\mathrm{Ai}(s)+\beta\,\mathrm{Bi}(s)\bigr]},\\
		y &= \frac{1}{\bigl[\alpha\,\mathrm{Ai}(s)+\beta\,\mathrm{Bi}(s)\bigr]^2},
	\end{aligned}
\end{equation}
where $s\in[s_1,\infty)$, and $s_1$ is defined by
\begin{equation}
	\alpha\,\mathrm{Ai}(s_1)+\beta\,\mathrm{Bi}(s_1)=0.
\end{equation}
Here, $\mathrm{Ai}$ and $\mathrm{Bi}$ are Airy functions. The contact-line position $\xi\to0$ corresponds to $s\to\infty$. The coefficient $\beta$ is determined by matching \eqref{eq:paraSol} to the full solution of \eqref{eq:H}, which yields
\begin{equation}\label{eq:beta2}
	\beta^{2}=\pi\,\exp\!\left(-\frac{1}{3\delta}\right)/2^{2/3}+O(\delta).
\end{equation}

Equations \eqref{eq:simply}--\eqref{eq:beta2} follow the asymptotic analysis of \citet{snoeijer2010asymptotic} for the rim profile of a dewetting film. We now examine the large-$s$ limit corresponding to the near-contact-line region. For $s\to\infty$, the Airy functions admit the following expansions
\begin{equation}\label{eq:large-s}
	\begin{aligned}
		\mathrm{Ai}(s) &= \frac{e^{-(2/3)s^{3/2}}}{2\sqrt{\pi}\,s^{1/4}}
		\left(1-\frac{5}{48}s^{-3/2}+O(s^{-3})\right),\\
		\mathrm{Bi}(s) &= \frac{e^{(2/3)s^{3/2}}}{\sqrt{\pi}\,s^{1/4}}
		\left(1+\frac{5}{48}s^{-3/2}+O(s^{-3})\right).
	\end{aligned}
\end{equation}
Substituting \eqref{eq:large-s} into \eqref{eq:paraSol} gives
\begin{equation}\label{eq:sol-large-s}
	\begin{aligned}
		\xi &=
		\frac{2^{1/3}\pi}
		{\beta^2\bigl[\alpha/\beta+\mathrm{Bi}(s)/\mathrm{Ai}(s)\bigr]}
		\approx
		\frac{2^{1/3}\pi}{2\beta^2}
		\exp\!\left(-\frac{4}{3}s^{3/2}\right),\\
		y &=
		\frac{1}
		{\beta^2\mathrm{Bi}(s)^2
			\bigl[1+(\alpha/\beta)\,\mathrm{Ai}(s)/\mathrm{Bi}(s)\bigr]^2}
		\approx
		\frac{\pi s^{1/2}}{\beta^2}
		\exp\!\left(-\frac{4}{3}s^{3/2}\right).
	\end{aligned}
\end{equation}
In obtaining \eqref{eq:sol-large-s}, the condition $\alpha/\beta\ll \mathrm{Bi}(s)/\mathrm{Ai}(s)$ is used. Based on the discussion by \citet{snoeijer2010asymptotic}, $\alpha/\beta \sim Ca^{1/3}w/h_0$ is small compared with $\mathrm{Bi}(s)/\mathrm{Ai}(s)\approx 2 \exp[(4/3)s^{3/2}]$ as $s\to\infty$.

%\begin{figure}[t]
%	\centering
%	\begin{subfigure}[t]{0.48\textwidth}
%		\centering
%		\includegraphics[width=\linewidth]{figs/fig-A-2a.pdf}
%		\subcaption{}
%	\end{subfigure}\hfill
%	\begin{subfigure}[t]{0.48\textwidth}
%		\centering
%		\includegraphics[width=\linewidth]{figs/fig-A-2b.pdf}
%		\subcaption{}
%	\end{subfigure}
%	
%	\medskip % or \vspace{0.5em}
%	
%	\begin{subfigure}[t]{0.48\textwidth}
%		\centering
%		\includegraphics[width=\linewidth]{figs/fig-A-2c.pdf}
%		\subcaption{}
%	\end{subfigure}\hfill
%	\begin{subfigure}[t]{0.48\textwidth}
%		\centering
%		\includegraphics[width=\linewidth]{figs/fig-A-2d.pdf}
%		\subcaption{}
%	\end{subfigure}
%	\caption{Interface slope versus distance from the contact line at different times for $\theta_e = 5^\circ$ and $\Delta T = 4\,\mathrm{K}$. (a) $f_{ac}=0.05$, (b) $f_{ac}=0.1$, (c) $f_{ac}=0.2$, and (d) $f_{ac}=0.4$.}
%	\label{fig:A-2}
%\end{figure}

Combining \eqref{eq:beta2} and \eqref{eq:sol-large-s}, we obtain
\begin{equation}
	y(\xi)=2^{2/3}\left\{-\frac{3}{4}\ln\!\left[e^{-1/(3\delta)}\xi\right]\right\}^{1/3}\xi.
\end{equation}
Differentiating with respect to $\xi$ gives
\begin{equation}\label{eq:rim-sol}
	y'(\xi)=2^{2/3}u(\xi)^{1/3}\left[1-\frac{1}{4u(\xi)}\right],
\end{equation}
where $u(\xi)=1/(4\delta) - (3/4)\ln\xi$.

Using the rescaling relation
\begin{equation}
	\frac{\mathrm{d}h}{\mathrm{d}x}
	= \theta_e\,\delta^{1/3}\,\frac{\mathrm{d}y}{\mathrm{d}\xi},
\end{equation}
equation \eqref{eq:rim-sol} can be rewritten in dimensional form as
\begin{equation}\label{eq:sol-corr}
	h'(x)^3 = \left( \theta_e^3 - 9\,Ca\ln\xi \right)
	\left(1 - \frac{3\,Ca}{ \theta_e^3 - 9\,Ca\ln\xi}\right)^3.
\end{equation}

\section{Interfacial-slope distribution}\label{appC}

\begin{figure}[t]
	\centering
	\begin{subfigure}[t]{0.48\textwidth}
		\centering
		\includegraphics[width=\linewidth]{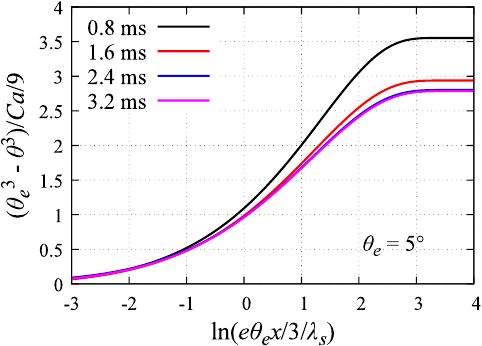}
		\subcaption{}
	\end{subfigure}\hfill
	\begin{subfigure}[t]{0.48\textwidth}
		\centering
		\includegraphics[width=\linewidth]{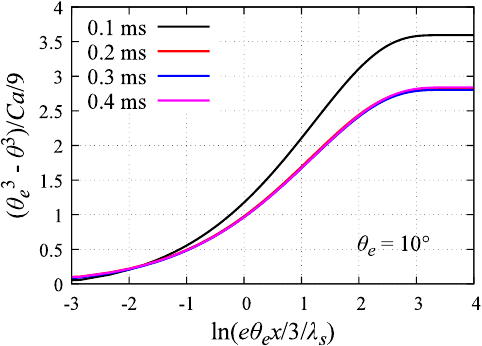}
		\subcaption{}
	\end{subfigure}
	
	\medskip % or \vspace{0.5em}
	
	\begin{subfigure}[t]{0.48\textwidth}
		\centering
		\includegraphics[width=\linewidth]{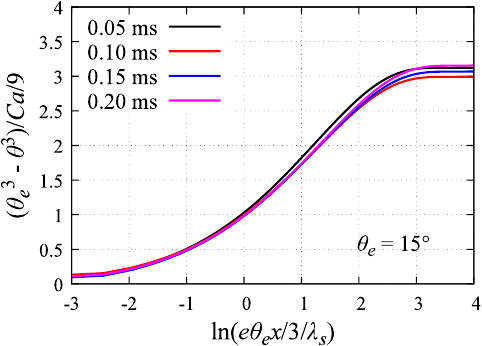}
		\subcaption{}
	\end{subfigure}\hfill
	\begin{subfigure}[t]{0.48\textwidth}
		\centering
		\includegraphics[width=\linewidth]{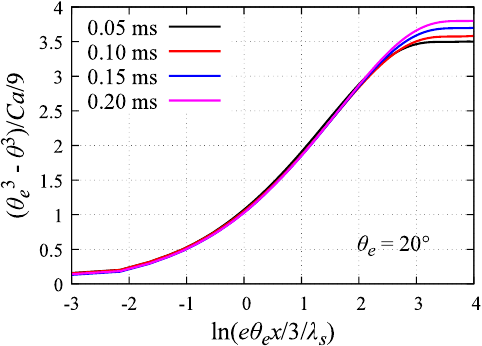}
		\subcaption{}
	\end{subfigure}
	\caption{Temporal evolution of interfacial-slope distributions for isothermal dewetting, plotted using scaling \eqref{solution0K-lin}. Results are shown at different times for (a) $\theta_e=5^{\circ}$, (b) $\theta_e=10^{\circ}$, (c) $\theta_e=15^{\circ}$ and (d) $\theta_e=20^{\circ}$.}
	\label{fig:A-1}
\end{figure}

\begin{figure}[t]	
	\begin{subfigure}[t]{0.48\textwidth}
		\centering
		\includegraphics[width=\linewidth]{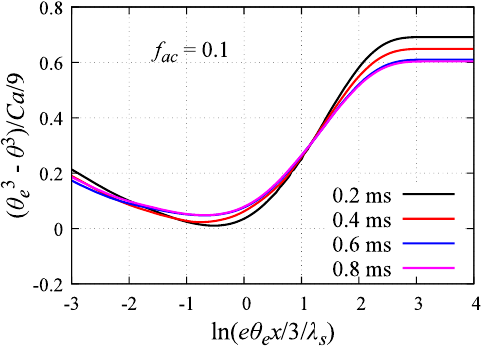}
		\subcaption{}
	\end{subfigure}\hfill
	\begin{subfigure}[t]{0.48\textwidth}
		\centering
		\includegraphics[width=\linewidth]{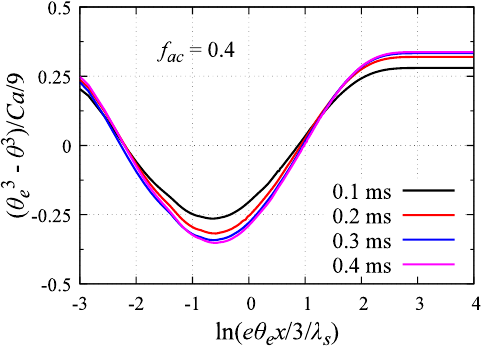}
		\subcaption{}
	\end{subfigure}
	\caption{Temporal evolution of the interfacial-slope distributions for dewetting with phase change, plotted using the Cox--Voinov scaling implied by \eqref{eq:zhang}, at $\theta_e=5^{\circ}$ and $\Delta T=8\,\mathrm{K}$. (a) $f_{ac}=0.1$. (b) $f_{ac}=0.4$.}
	\label{fig:A-4}
\end{figure}

\begin{figure}[t]
	\centering
	\includegraphics[width=0.5\linewidth]{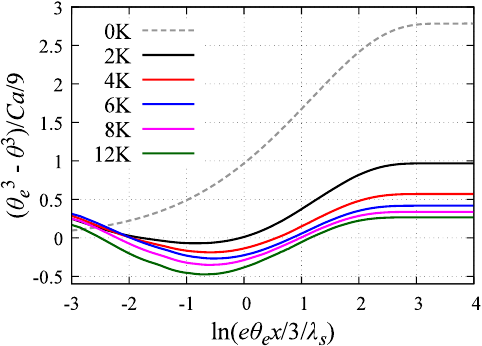}
	\caption{Interfacial-slope distributions at different wall superheats, replotted using the Cox--Voinov scaling implied by \eqref{eq:zhang}, for $\theta_e=5^{\circ}$ and $f_{ac}=0.4$. The profiles are taken after the slope distribution has reached a quasi-steady state. The dashed curves denote the isothermal-dewetting result at $\theta_e=5^{\circ}$.}
	\label{fig:A-5}
\end{figure}

To assess the quasi-steady assumption underlying the analysis of the interfacial-slope distribution, figures~\ref{fig:A-1} and \ref{fig:A-4} show its temporal evolution from the early transient stage to later times for isothermal dewetting and dewetting with phase change, respectively. Figure~\ref{fig:A-1} presents the results for four equilibrium contact angles, whereas figure~\ref{fig:A-4} considers representative evaporating cases at $\theta_e=5^\circ$ and $\Delta T=8\,\mathrm{K}$ for $f_{ac}=0.1$ and $0.4$. At the earliest times, the rescaled slope profiles differ visibly from those obtained at later times, reflecting the initial acceleration of the contact line and the relaxation of the interface from the prescribed initial profile. As time progresses, the temporal variation gradually weakens and successive profiles approach one another. The two latest profiles nearly overlap over most of the resolved range, with only minor differences remaining at larger distances from the contact line. This behaviour is observed for both isothermal and evaporating cases, despite the more complex slope variation induced by phase change in the latter case. These results indicate that, in a frame moving with the contact line, the interfacial-slope distribution progressively approaches a quasi-steady form. They therefore support the use of representative time instants in the slope analyses of \S\ref{sec:3.1} and \S\ref{sec:3.2}, as well as the quasi-steady approximation adopted in the lubrication analysis.

%Figure~\ref{fig:A-5} complements figure~\ref{fig:3-2-5} by extending the assessment of the Cox--Voinov-type relation \eqref{eq:zhang} to $f_{ac}=0.4$ at $\theta_e=5^\circ$. Although the influence of phase change is weaker in this case, the curves obtained at different wall superheats still exhibit systematically different slopes in the logarithmic representation. This indicates that the Cox--Voinov-type relation \eqref{eq:zhang} alone does not fully collapse the interfacial-slope distributions, even in the weak-phase-change regime. When considered together with figure~\ref{fig:3-2-6}(a), where inclusion of the first-order evaporation correction makes the rescaled curves nearly parallel to the isothermal result in the intermediate region, this observation provides further support for the evaporation-corrected expression \eqref{solution-large-x}.

Figure~\ref{fig:A-5} complements figure~\ref{fig:3-2-5} by extending the comparison with the Cox--Voinov-type relation \eqref{eq:zhang} to $f_{ac}=0.4$ at $\theta_e=5^\circ$. In this case, the parameter $\varepsilon$ defined in \eqref{eq:ratio} ranges from about 2.5 to 4, so that evaporation dominates near the contact line and the interfacial slope exceeds $\theta_e$ there, giving negative minima in the rescaled curves. Beyond these minima, however, the curves remain considerably flatter than the isothermal curve. This is consistent with $R(h_0/l_R)\approx0.3$--$0.5$, comparable to that for the cases in figure~\ref{fig:3-2-5}. In other words, the evaporation contribution outside the micro-region remains significant, so that a Cox--Voinov region does not develop.

\section{Local logarithmic variation of the interfacial slope with evaporation}
\label{app:local_log_slope}

In this appendix, we derive approximate logarithmic relations for the interfacial slope in different regions near the contact line with evaporation. The derivation is based on the quasi-steady lubrication equation introduced in \S \ref{sec:slope}, and is intended to describe the local slope variation over a finite spatial interval rather than to establish a uniformly valid matched-asymptotic solution.

The dimensional form of the lubrication equation \eqref{lubrication} can be written as
\begin{equation}
	\frac{\mathrm{d}}{\mathrm{d}x}
	\left[
	h^2\left(\lambda_s+\frac{h}{3}\right)
	\frac{\mathrm{d}^3h}{\mathrm{d}x^3}
	-Ca\,h
	\right]
	+
	Ca_e\frac{\l_R}{h+\l_R}=0.
	\label{eq:app_lubrication}
\end{equation}

Integrating \eqref{eq:app_lubrication} once and taking the integration constant to be zero, which corresponds to the vanishing liquid flux at the contact line, yields
\begin{equation}
	h^2\left(\lambda_s+\frac{h}{3}\right)
	\frac{\mathrm{d}^3h}{\mathrm{d}x^3}
	=
	Ca\,h
	-
	Ca_e\mathcal{F}_e(x)\,x,
	\label{eq:app_integrated}
\end{equation}
where $\mathcal{F}_e(x)$ is defined as 
\begin{equation}
	\mathcal{F}_e(x)
	=
	\frac{1}{x}\int_0^x
	\frac{\l_R}{h(\xi)+\l_R}\,\mathrm{d}\xi.
	\label{eq:app_evap_factor}
\end{equation}

Following the logarithmic-coordinate transformation employed by  \citet{hocking1995contact}, we introduce
\begin{equation}
	s(\tau)={h(x)}/{x}, \qquad \tau=\ln\left({x/l_R}\right).
	\label{eq:app_secant_slope}
\end{equation}
where $s(\tau)$ is the secant slope of the interface measured from the contact line. With this rescaling, \eqref{eq:app_integrated} can be rewritten as
\begin{equation}
	s_\tau-s_{\tau\tau\tau}
	=
	\frac{3}{s^3}
	\frac{h}{h+3\lambda_s}
	\left[Ca_e{\mathcal{F}_e(x)}-Ca\,s\right].
	\label{eq:app_secant_equation}
\end{equation}

The secant slope $s$ is related to the local interfacial slope $\theta$ through
\begin{equation}
	\theta(x)=\frac{\mathrm{d}h}{\mathrm{d}x}
	=s+s_\tau.
	\label{eq:app_tangent_slope}
\end{equation}
It follows that
\begin{equation}
	\theta_\tau-\theta_{\tau\tau}
	=s_\tau-s_{\tau\tau\tau}.
	\label{eq:app_tangent_identity}
\end{equation}

%This factor accounts for the reduction in evaporation rate with increasing film thickness. Since $0<\mathcal{F}_e\leq 1$, it approaches unity when $h\ll\ell_R$ throughout the interval between the contact line and the position considered. Equation \eqref{eq:app_integrated} then becomes

Combining \eqref{eq:app_secant_equation} and \eqref{eq:app_tangent_identity}, and multiplying by $4\theta^3$, gives
\begin{equation}
		\frac{\mathrm{d}\theta^4}{\mathrm{d}\tau}
		-
		4\theta^3\theta_{\tau\tau}
		=
		12\left(\frac{\theta}{s}\right)^3
		\frac{h}{h+3\lambda_s}
		\left[Ca_e\mathcal{F}_e(x)-Ca\,s\right].
	\label{eq:app_theta4_exact}
\end{equation}

Using the slowly varying slope approximation, we assume that $|\theta_{\tau\tau}|\ll|\theta_\tau|$, so that the second term on the left-hand side of \eqref{eq:app_theta4_exact} can be neglected, yielding
\begin{equation}
	\frac{\mathrm{d}\theta^4}{\mathrm{d}\ln x}
	\approx
	12\left(\frac{\theta}{s}\right)^3
	\frac{h}{h+3\lambda_s}
	\left[Ca_e\mathcal{F}_e(x)-Ca\,s\right].
	\label{eq:app_theta4_local}
\end{equation}
For a stationary contact line ($Ca=0$), assuming $3\lambda_s\ll h\ll\l_R$ and $\theta\approx s$, the right-hand side of \eqref{eq:app_theta4_local} reduces to $12Ca_e$, recovering the logarithmic relation obtained by \citet{hocking1995contact}. Contact-line recession introduces an opposing contribution, thereby reducing the logarithmic coefficient associated with evaporation.

Similarly, combining \eqref{eq:app_secant_equation} and \eqref{eq:app_tangent_identity}, and multiplying by $3\theta^2$, gives
\begin{equation}
	\frac{\mathrm{d}\theta^3}{\mathrm{d}\tau}
	-
	3\theta^2\theta_{\tau\tau}
	=
	9\left(\frac{\theta}{s}\right)^2
	\frac{h}{h+3\lambda_s}
	\left[
	\frac{Ca_e\mathcal{F}_e(x)}{s}-Ca
	\right].
	\label{eq:app_theta3_exact}
\end{equation}

Neglecting the second-order derivative term under the same slowly varying slope approximation, we obtain
\begin{equation}
	\frac{\mathrm{d}\theta^3}{\mathrm{d}\ln x}
	\approx
	9\left(\frac{\theta}{s}\right)^2
	\frac{h}{h+3\lambda_s}
	\left[
	\frac{Ca_e\mathcal{F}_e(x)}{s}-Ca
	\right].
	\label{eq:app_theta3_local}
\end{equation}
In the absence of evaporation ($Ca_e=0$), assuming $3\lambda_s\ll h$ and $\theta\approx s$, the right-hand side of \eqref{eq:app_theta3_local} reduces to $-9Ca$, recovering the classical Cox--Voinov relation for a receding contact line. Conversely, evaporation introduces a positive contribution, reducing the magnitude of the negative logarithmic coefficient induced by contact-line recession.

To estimate the logarithmic coefficients from the DNS results, we consider the two spatial intervals in which $\theta^4$ and $\theta^3$ exhibit approximately linear variations with $\ln x$, as shown in figure \ref{fig:3-5-3}.

For the near-contact-line region in which $\theta^4$ varies approximately linearly with $\ln x$, we assume $h\ll\l_R$, so that $\mathcal{F}_e(x)\approx1$. Equation \eqref{eq:app_theta4_local} then reduces to
\begin{equation}
	\frac{\mathrm{d}\theta^4}{\mathrm{d}\ln x} \approx
	B_4(x) =
	12\left(\frac{\theta}{s}\right)^3
	\frac{h}{h+3\lambda_s}
	\left(Ca_e-Ca\,s\right).
	\label{eq:app_B4_local}
\end{equation}

Farther from the contact line in which $\theta^3$ varies approximately linearly with $\ln x$, we assume $h\gg3\lambda_s$ and $h\gg\l_R$. To estimate $\mathcal{F}_e(x)$, we approximate the interface between the contact line and a given position $x$ by a straight wedge, $h(\xi)\simeq s(x)\xi$. Substituting this approximation into \eqref{eq:app_evap_factor} gives
\begin{equation}
	\begin{aligned}
		\mathcal{F}_e(x)
		\approx
		\frac{\l_R}{h}
		\ln\left(\frac{h}{\l_R}\right).
	\end{aligned}
	\label{eq:app_Fe_wedge}
\end{equation}
Using this approximation, \eqref{eq:app_theta3_local} reduces to
\begin{equation}
	\frac{\mathrm{d}\theta^3}{\mathrm{d}\ln x}
	\approx B_3(x)
	=
	9\left(\frac{\theta}{s}\right)^2
	\left[
	-Ca+
	\frac{Ca_e}{s}
	\frac{\l_R}{h}
	\ln\left(\frac{h}{\l_R}\right)
	\right].
	\label{eq:app_B3_local}
\end{equation}

The logarithmic coefficients are then estimated by averaging $B_4(x)$ and $B_3(x)$ over
their respective fitting intervals,
\begin{equation}
	\overline{B}_p=
	\frac{1}{\ln(x_2/x_1)}
	\int_{x_1}^{x_2}
	B_p(x)\,\frac{\mathrm{d}x}{x},
	\qquad p=3,4.
	\label{eq:app_B_average}
\end{equation}
The local film thickness $h(x)$, tangent slope $\theta(x)$ and secant slope $s(x)=h(x)/x$ used in this evaluation are obtained from the DNS results.

As shown in figure~\ref{fig:3-5-3}, $\theta^4$ exhibits an approximately linear variation with $\ln x$ over $2.14\times10^{-7}\leq x\leq6.41\times10^{-7}\,\mathrm{m}$. Numerical integration of \eqref{eq:app_B_average} yields $\overline{B}_4=2.23\times10^{-4}$, which agrees reasonably well with the DNS-fitted value of $2.09\times10^{-4}$. For the $\theta^3$ relation, numerical integration over $3.62\times10^{-6}\leq x\leq2.25\times10^{-5}\,\mathrm{m}$ gives $\overline{B}_3=-7.48\times10^{-4}$, compared with the DNS-fitted value of $-1.31\times10^{-3}$, indicating an underestimation of the magnitude of the logarithmic slope.

\end{appen}

\bibliographystyle{jfm}
\bibliography{jfm}

%% End of file `jfm.bib'.

\end{document}